\documentclass[12pt]{article}
\usepackage{amsmath,multirow}
\usepackage{amssymb}
\usepackage{graphicx}
\usepackage[comma]{natbib}
\usepackage[colorlinks,linkcolor=red,anchorcolor=blue,citecolor=blue]{hyperref}
\usepackage{float}
\usepackage{bm}
\usepackage{longtable}
\usepackage{multirow,booktabs}
\usepackage{lineno,hyperref}
\usepackage{rotating}
\usepackage{chngpage}
\usepackage{color, soul}
\usepackage{diagbox}
\usepackage{lscape}
\usepackage{enumerate}
\usepackage{dsfont}
\usepackage[justification=centering]{caption}
\newtheorem{theorem}{Theorem}

\newtheorem{remark}{Remark}

\newtheorem{definition}{Definition}
\newtheorem{lemma}{Lemma}

\makeatletter \def\@biblabel#1{#1.} \makeatother
\allowdisplaybreaks

\catcode`@=11 \@addtoreset{equation}{section} \catcode`@=12

\begin{document}
\title{Change-points analysis for generalized integer-valued autoregressive model via minimum description length principle
\footnotetext{$^1$School of Mathematics and Statistics, Liaoning University, Shenyang, China\\
\indent~~$\ast$Corresponding author, E-mail: wangdehui@lnu.edu.cn\\
}
}
\author{Danshu Sheng$^1$, Dehui Wang$^{1*}$}
\date{}
\maketitle
\begin{center}
\begin{minipage}{14.5truecm}
{{\bf Abstract}~~This article considers the problem of modeling a class of nonstationary count time series using multiple change-points generalized integer-valued autoregressive (MCP-GINAR) processes.
The minimum description length principle (MDL) is applied to study the statistical inference for the MCP-GINAR model,
and the consistency results of the MDL model selection procedure are established respectively under the condition of known and unknown number of change-points.
To find the ``best" combination of the number of change-points, the locations of change-points, the order of each segment and its parameters, a genetic algorithm with simulated annealing is implemented to solve this difficult optimization problem.
In particular, the simulated annealing process makes up for the precocious problem of the traditional genetic algorithm.
Numerical results from simulation experiments and three examples of real data analyses show that the procedure has excellent empirical properties.}
\\
~\\
$\mathbf{Keywords:}$ GINAR($p$) model $\cdot$ multiple change-points $\cdot$  minimum description length $\cdot$ genetic
algorithm $\cdot$  simulated annealing
\end{minipage}
\end{center}
\section{Introduction}
Modeling and analysis of count time series  have attracted a lot of attention over the last years. Since \cite{Al1987} proposed the first order integer-valued autoregressive (INAR(1)) time series model based on binomial thinning operator (\cite{SV1979}), modeling INAR-type models based on different thinning operators have become a common approach and have been widely used in many fields like epidemiology, social sciences, economics, life sciences and others.
Furthermore, to make the integer-valued models based on thinning more flexible for practical purposes,
\cite{Latour1998} generalized the binomial thinning operator to a generalized thinning operator and proposed the following causal and stationary generalized integer-valued autoregressive (GINAR($p$)) process,
\begin{definition}\label{GINAR}
The GINAR($p$) model $\{X_{t}\}_{t\geqslant1}$ is defined by the following recursion
\vspace{-1mm}
\begin{align}\label{ginar}
X_{t}=\sum\limits_{k=1}^p\alpha_{k}\circ_G X_{t-k}+Z_{t},
\end{align}
where $\alpha\circ_G X=\sum\limits_{i=1}^{X}B_i$, $\{B_i\}$ is an independent identically distributed (i.i.d.) integer-valued random sequence with mean $\alpha\in(0,1)$ and independent of non-negative integer-valued random variable $X$. $\{Z_{t}\}$ is a sequence of i.i.d. integer-valued random variables with mean $\gamma$ and $Z_{t}$ is not depending on past values of $\{X_{s}\}_{s<t}$.
\end{definition}
Due to the flexibility and practicability of the GINAR models, a large quantity of articles focusing on the modeling and statistical inference for such models have arisen.
For example, \cite{Ristic2009}, \cite{Nastic2012}, \cite{Scotto2018}, \cite{BS2019}, and \cite{Darolles2019} considered the modeling of the INAR-type models to better handle the fitting problems.
\cite{Zheng2006}, \cite{SilvaSilva2006}, \cite{Bu2008} studied the parameter estimation for the GINAR($p$) models.
\cite{F2016}, \cite{F2021}, \cite{Henderson2018}, \cite{Guan2022} and \cite{Gourieroux2004} applied the INAR-type models to under-reported data, longitudinal data and insurance actuarial.

Note that the GINAR models are typically used to describe stationary processes, while it is well known that non-stationary processes are also a common phenomenon in statistics.
Although complex non-stationary models have been developed in different fields, they are often difficult to explain.
The concept of piecewise stationary models has become a popular method by dividing nonstationary data into several stationary parts.
Among the different types of piecewise stationary models, the so-called multiple change-points (MCP) models have received special attention.
Studies of change-points models date back to \cite{Page1954,Page1955}. Since then, this topic has been of interest to statisticians and researchers in many other fields. Such as, in financial analysis, \cite{Russell2019} identified inflation regime by detecting changes in the mean of a stationary process, in medicine, \cite{Jong2003} identified the presence of tumors by detecting breaks in the Array Genome sequence, in climatology, \cite{Shi2022} analyzed the causes of temperature changes by detecting structural changes in the Central England temperature series.

Based on different data types, the study of time series can be divided into continuous and count time series models.
So far, many excellent articles, such as \cite{Lee2003}, \cite{Davis2006}, \cite{Chan2014}, \cite{Chen2021}, \cite{Aue2013}, \cite{Niu2016}, \cite{Casini2018}, \cite{Truong2020}, just to name a few, have studied and reviewed methodological issues related to estimation, detection and computation for continuous time series models involving structural changes.
In contrast, the research of count time series models involving structural changes is still mainly focused on the change-points detection. Such as, among others,
\cite{Kang2009}, \cite{YuandKim2020} and \cite{Lee2022} considered the problem of testing for a parameter change in different types of INAR(1) models by taking advantage of the cumulative sum (CUSUM) test. \cite{DiopKengne2021}, \cite{DiopKengne2022} constructed test statistics, which is based on Poisson quasi-maximum likelihood estimator of the parameters, to detect changes in the mean parameters of the generalized integer-valued models, and then used this procedure to detect changes in the number of epidemic cases.
\cite{Chattopadhyay2021} considered the problem of change-point analysis for the INAR(1) model with time-varying covariates.
\cite{Yu2022} applied the empirical likelihood ratio (ELR) test to uncover a structural change in INAR processes.

However, there is surprisingly little research on the change-points estimation of count time series.
To task this, the main goal of this article is to introduce the following piecewise stationary model, multiple change-points generalized integer-valued autoregressive model (MCP-GINAR), to describe the common non-stationary processes in count time series, and then study its change-points estimation,
\begin{definition}
The MCP-GINAR process with $m$ change-points $\{X_t\}_{t=1}^{n}$ is defined by the recursion:
\begin{align}
X_{t}
=\begin{cases}
\alpha_{1,1}\circ_G X_{t-1,1}+...+\alpha_{p_1,1}\circ_G X_{t-p_1,1}+Z_{t,1},~0<t\leq\tau_{1},\\
\vdots\\
\alpha_{1,j}\circ_G X_{t-1,j}+...+\alpha_{p_j,j}\circ_G X_{t-p_j,j}+Z_{t,j},~\tau_{j-1}<t\leq\tau_{j},\\
\vdots\\
\alpha_{1,m}\circ_G X_{t-1,m}+...+\alpha_{p_m,m}\circ_G X_{t-p_m,m}+Z_{t,m},\tau_{m}<t\leq n,\\
\end{cases}\label{CGINAR}
\end{align}
where $\bm{\tau}=(\tau_1,\tau_2,...,\tau_{m})$ denotes the vector of unknown locations of change-points, $\tau_0=0$ and $\tau_{m+1}=n$.
Each change-points location $\tau_j$ is an integer between $1$ and $n-1$ inclusive and the change-points are ordered such that $\tau_{j_1}< \tau_{j_2}$ if, and only if, $j_1<j_2$.
The time series $\bm{x}_n=(X_1, X_2,..., X_n)$ can be written as \begin{align}\label{xn}
\bm{x}_n=(X_{1,1},...,X_{n_1,1},X_{1,2},...,X_{n_2,2},...,X_{1,j},...,X_{n_j,j},...,X_{1,m+1},...,X_{n_{m+1},m+1}),
\end{align}
where $n_j=\tau_j-\tau_{j-1}$ for $j = 1,...,m+1$ and $n = n_1 + n_2 +...+ n_{m+1}$.
In particular, the $j$-th piece $\{X_{t,j}\}_{t=1}^{n_j}$ of the series is modeled as an GINAR process (\ref{ginar}) with order $p_j$,
\begin{align}\label{GINARj}
X_{t,j}=\alpha_{1,j}\circ_G X_{t-1,j}+...+\alpha_{p_j,j}\circ_G X_{t-p_j,j}+Z_{t,j},
\end{align}
where ``$\circ_G$" is defined in Definition \ref{GINAR}, $\{Z_{t,j}\}$ is a sequence of i.i.d. random variables with mean $\gamma_j>0$. $\bm{\theta_j}=(\alpha_{1,j},...,\alpha_{p_j,j},\gamma_j)$ is the parameter vector corresponding to this GINAR($p_j$) process, which is assumed to be an interior point of the compact space $\Theta_j(p_j)=(0,+\infty)\times(0,1)^{p_j}$.
\end{definition}

For change-points estimation, we can regard it as a model selection problem.
One common approach is to minimize a specific information criterion (IC) to select an optimal model.
Given that our main focus is on estimating unknown parameters (the change-points number $m$, the change-points locations $\bm{\tau}=(\tau_1,\tau_2,...,\tau_{m})$, the order of each segment GINAR process $p_j$ and the corresponding parameter $\bm{\theta_j}=(\alpha_{1,j},...,\alpha_{p_j,j},\gamma_j)$) in the model (\ref{CGINAR}), we apply the minimum description length (MDL) principle of \cite{Rissanen1989} to define a best-fitting model.
MDL principle is designed to select a model that enables maximum compression of the data as the best-fitting model.
Since \cite{Davis2006} proposed a specific MDL applied to change-points estimation,
this criterion has been studied by many scholars and applied to various model selection problems.
For example,
\cite{DavisYau2013}, \cite{Davis2016} discuss the consistency of MDL model selection procedure,
\cite{Aue2014} proposed a criterion based on MDL principle that could solve both variable selection and change-points estimation,
\cite{Mhalla2020}, \cite{Woody2021}, \cite{Shi2022}, among others, success in applying MDL criterion to a variety of practical applications.

Furthermore, it is worth mentioning that since the search space (consisting of $m$, $\tau_j$, $p_j$ and $\bm{\theta_j}$) is huge, practical optimization of this criterion is not a trivial task. \cite{Davis2006}, \cite{Aue2014}, \cite{Doerr2017}, among others, have pointed out that genetic algorithm is an effective way to solve this difficulty.
However, although the genetic algorithm constructed by them has the characteristics of fast search speed, strong randomness, simple process and strong flexibility, it does not take into account the precocious characteristics of genetic algorithm.
Prematurity is a common problem of many global optimization algorithms, that is, the algorithm loses the population diversity prematurely, and then converges to the local optimal solution.
For genetic algorithm, it is difficult to introduce new genes through selection and crossover operation, and only mutation operation can make population transfer.
Thus, GA will remain in the old state for a very long time if the probability of mutation is small.
Search is ineffective and precocious may eventually result.
To effectively explore a better solution,
a locally jittering the chromosome method, the simulated annealing process,
is used to modify the classical GA \citep{Davis2006} in this paper.
The details are described in Section \ref{4.6}.

The main contributions of this article are as follows.
First, we propose the MCP-GINAR process to model a class of nonstationary count time series, and solve the multiple change-points estimation problem by using the MDL criterion. It includes the number of change-points, the position of change-points, the order of each segment and its parameters.
Second, we discuss respectively the consistency theorems of parameters based on the two case, the number of change-points known and unknown.
Third, we propose an automatic piecewise generalized integer autoregressive modeling program with simulated annealing (Auto-PGINARM-SA) to solve the optimizing problems of MDL criterion.
In particular, the simulated annealing process makes up for the precocious problem of the traditional genetic algorithm.

The rest contents of this article are organized as follows.
In Section 2, we state some statistical properties and inferences of GINAR($p$) model.
In Section 3, we set some notations, derive the estimation procedure and provide the consistency of parameters.
Section 4 describes the implementation process of Auto-PGINARM-SA procedure.
In Section 5, we study the performance of the Auto-PGINARM-SA via simulation, and apply the procedure to three applications in Section 6.
The article ends with a conclusion section and all proofs are given in Appendix.

\section{Statistical Inferences for GINAR($p$) model}\label{Set2}
In this section, we state some statistical properties and inferences of the $j$-th piece GINAR($p_j$) model.
Considering the strict stationarity and ergodicity of the GINAR($p_j$) process and the fact that GINAR($p_j$) can be written as the AR($p_j$) process will be helpful for us to establish the consistency of the parameter estimators. Therefore, we condense the conclusions verified by scholars into the following two Remarks.
\begin{remark}\label{remark1}
If $0<\sum\limits_{i=1}^{p_j}\alpha_{i,j}<1$, then there exists a unique strictly stationary and ergodic $\{X_{t,j}\}_{t=1}^{n_j}$ that satisfies(\ref{GINARj}) (\cite{Zhang 2010}). Additionally, if each segment is strictly stationary and ergodic GINAR process, we refer to the MCP-GINAR process (\ref{CGINAR}) as a piecewise stationary process.
\end{remark}

\begin{remark}\label{remark2}
Let $\{X_{t,j}\}_{t=1}^{n_j}$ be a GINAR($p_j$) process satisfying (\ref{GINARj}). Then $\{X_{t,j}\}_{t=1}^{n_j}$ is an AR($p_j$) that can be written as
\begin{align}\label{AR(pj)}
e_{t,j}=\sum\limits_{k=0}^{p_j}\alpha_{k,j}( X_{t-k,j}-\mu_{X_j}),
\end{align}
where $\alpha_{0,j}=1$, $\mu_{X_j}$ is the mean of the $j$-th piece $\{X_{t,j}\}_{t=1}^{n_j}$,
$e_{t,j}$ is a white noise process of variance $\sigma_{e_{j}}^2=\mu_{X_j}\sum\limits_{k=1}^{p_j}\beta_{k,j}+\sigma_{Z_j}^2$, $\beta_{k,j}$ and $\sigma_{Z_j}^2$ are the variance of the variables $B_{k,j}$ and $Z_{t,j}$ respectively (\cite{Latour1998}).
\end{remark}
\subsection{Poisson quasi-maximum likelihood}
The MDL can be regarded as the negative of the sum of the log-likelihood for each of the segments plus a penalty term which penalizes the size of the model. In this study, the Poisson quasi-maximum likelihood (PQML) is used as the cost function of MDL criterion.
On the one hand, if  the maximum likelihood (ML) proposed by \cite{Bu2008} is used as the cost function,
since the transition probabilities in ML are written using the convolution formula, the form is too complex, it is challenging to confirm the corresponding conditions when discussing the consistency of the number of change-points and the location of change-points.
On the other hand, as mentioned in \cite{Ahmad Francq2016}, it is known that certain ML can be consistent and asymptotically normal for the parameters of the conditional mean and variance, even if the actual conditional distribution is not the one which is assumed by the ML. Noting that the parameters in GINAR($p$) can be treated as mean parameters and that PQML can be shown to have consistency and asymptotic normality under certain assumptions, so the PQML is a great option as the cost function.

Given all the past observations, the conditional Poisson quasi-maximum log-likelihood of the $j$-th piece, ${\bm{x_j}}=\{X_{t,j}\}_{t=-p}^{n_j}$, is obtained by
\begin{align}\label{PQMLeq}
L_{n_j}^{(j)}(\bm{\theta_j};p_j,\bm{x_j})&=\sum\limits_{t=1}^{n_j}\left[X_{t,j}\log \xi_{t,j}(\bm{\theta_j};p_j|X_{s,j},s<t)-\xi_{t,j}(\bm{\theta_j};p_j|X_{s,j},s<t)\right]\\
&=\sum\limits_{t=1}^{n_j}\ell_t^{(j)}(\bm{\theta_j};p_j, X_{t,j}|X_{s,j},s<t),\nonumber
\end{align}
where $\xi_{t,j}(\bm{\theta_j};p_j|X_{s,j},s<t)=\sum\limits_{k=1}^{p_j}\alpha_{k,j} X_{t-k,j}+\gamma_j$. It is important to note that $\bm{x_0}=\{X_{t,1}\}_{t=-p}^{-1}$ for the first piece, where $\bm{x_0}$ is a set of initial integer values. 
Then the PQML estimator of $\bm{\theta_j}$ is defined by
\begin{align*}
{\bm{\hat\theta}_j}=\arg\max\limits_{\bm{\theta_j}\in\Theta_j(p_j)} L_{n_j}^{(j)}(\bm{\theta_j};p_j,\bm{x_j}).
\end{align*}
Assuming that $0<\sum\limits_{i=1}^{p_j}\alpha_{p_i}<1$ and ${\rm E}(X_{1,j})<\infty$, as stated in Section 3.7. of \cite{Ahmad Francq2016}, ${\bm{\hat\theta}_j}$ satisfy the following asymptotic normality
\begin{align*}
\sqrt{n_j}(\bm{\hat{\theta}_j}-\bm{\theta_j}^0)\xrightarrow{d} N(0,\bm{\Sigma_j})~~~\text{as~~~$n_j\rightarrow \infty$},
\end{align*}
where $\bm{\theta_j}^0$ denote the true parameter value of $\bm{\theta_j}$. To simplify notation, denote $\xi_{t,j}(\bm{\theta_j})=\xi_{t,j}(\bm{\theta_j};p_j|X_{s,j},s<t)$, then the asymptotic variance matrix of the $j$-th piece PQML estimator can be consistently estimated by $\bm{\Sigma_j}=\bm{J_j^{-1}I_jJ_j^{-1}}$ with
\begin{align}
\bm{J_j}&=\frac{1}{n_j}\sum\limits_{t=1}^{n_j}\left.\frac{1}{\xi_{t,j}(\bm{\theta_j})}\frac{\partial\xi_{t,j}(\bm{\theta_j})}{\partial \bm{\theta_j}}\frac{\partial\xi_{t,j}(\bm{\theta_j})}{\partial \bm{\theta_j}^{\mathrm{T}}}\right|_{\bm{\theta_j}=\bm{\hat{\theta}_j}},\label{SE1}\\
\bm{I_j}&=\frac{1}{n_j}\sum\limits_{t=1}^{n_j}\left.(\frac{X_{t,j}}{\xi_{t,j}(\bm{\theta_j})}-1)^2\frac{\partial\xi_{t,j}(\bm{\theta_j})}{\partial \bm{\theta_j}}\frac{\partial\xi_{t,j}(\bm{\theta_j})}{\partial \bm{\theta_j}^{\mathrm{T}}}\right|_{\bm{\theta_j}=\bm{\hat{\theta}_j}}.\label{SE2}
\end{align}
\section{Main result}\label{Set3}
In this section, we first introduce the MDL criterion for this MCP-GINAR model based on PQML. And then we present the main results of the paper.
Theorem \ref{MDL_mknown_theorem} shows the strong consistency of the MDL procedure when the number of change-points is known.
Theorem \ref{week} not only shows the weak consistency of the MDL procedure when the number of change-points is unknown, but also gives the rate of convergence of the change-points estimators.
Furthermore, Theorems \ref{strong} gives the strongly consistency analog of Theorems \ref{week}, where stronger moment conditions are required.

For the sake of readability, we provide the following notations and assumption, and their corresponding explanations before introducing the MDL criterion.\\
~\\
$\emph{\textbf{Notations:}}$
\begin{itemize}
 \item Denote this whole class of MCP-GINAR models by $\mathcal{M}$ and any model from this class by $\mathcal{F} \in\mathcal{ M}$.
 \item Let ${\bm\lambda}=(\lambda_1,...,\lambda_{m})$, $0<\lambda_1<...<\lambda_{m}<1$, satisfy $\tau_j=[\lambda_jn]$, where $[x]$ is the greatest integer that is less than or equal to $x$.
 \item Upper bounds for GINAR orders $p$ are represented by $P_0$. Setting $\bm{p}=(p_1,...,p_{m+1})$ belongs to the parameter domain $\mathcal{P}=(0,P_0]^{m+1}\bigcap \mathds{Z}^{m+1}$, $\bm{\theta}=(\bm{\theta_1},...,\bm{\theta_{m+1}})$ belongs to the parameter domain $\bm{\Theta}=\prod\limits_{j=1}^{m+1}\bm{\Theta_j(p_j)}$.
 \item If the number and location of change points are known,
 the conditional Poisson quasi-maximum log-likelihood of the $j$-th piece can be given by (\ref{PQMLeq}).
However, in practice, both the number of change-points and the locations of the change-points are unknown. As a result,
for any observation $x_{i,j}$, its ``observed past", is in fact
$$(\bm{x_0},x_{1,1},...,x_{n_1,1},...,x_{1,j},...,x_{i-1,j}),$$ or equivalently, $\bm{\tilde{x}_{i,j}}\equiv(\bm{x_0},x_1,x_2,...,x_{\tau_{j-1}+i-1})$, where $\bm{x_0}=\{X_{t,1}\}_{t=-p}^{-1}$ is the initial value vector.
The conditional Poisson quasi log-likelihood of the $j$-th piece is then given by
\begin{align}\label{tilde_L}
\tilde{L}_{n_j}^{(j)}(\bm{\theta_j};p_j,\bm{x_j})&=\sum\limits_{t=1}^{n_j}x_{t,j}\log \xi_{t,j}(\bm{\theta_j};p_j|\bm{\tilde{x}_{i,j}})-\xi_{t,j}(\bm{\theta_j};p_j|\bm{\tilde{x}_{i,j}})\\
&=\sum\limits_{t=1}^{n_j}\ell_t^{(j)}(\bm{\theta_j};p_j,\bm{\tilde{x}_{i,j}}).\nonumber
\end{align}
\end{itemize}
$\emph{\textbf{Assumptions:}}$
\begin{itemize}
\item To ensure the piecewise stationary, for each segment, assume that $\bm{\theta}_j$ is an interior point of the compact space $\bm{\Theta}_j(p_j)$ and satisfies $\sum_{i=1}^{p_j}\alpha_{i,j}<1$, $\alpha_i\geq0$ for $i=1,...,p_j$.
 \item To accurately estimate the specified GINAR parameter values, the segments must have a sufficient number of observations.
If not, the estimation is overdetermined and the likelihood has an infinite value. So to ensure identifiability of the change-points, when we search for the change-points, we assume that there exists a $\epsilon_{\lambda}>0$ such that $\epsilon_{\lambda}<\min_{1\leq j\leq m}(|\lambda_j-\lambda_{j-1}|)$ and set
$$A_{\epsilon_{\lambda}}^m=\{{\bm\lambda}\in(0,1)^{m},0<\lambda_1<...<\lambda_{m}<1,\lambda_{j}-\lambda_{j-1}\geq\epsilon_{\lambda},j=1,...,m\}.$$
So under this restriction the number of change points is bounded by $M_0=[1/\epsilon_{\lambda}]+1$.
\end{itemize}
Let the vector $(m,\bm{\lambda},\bm{p},\bm{\theta})$ specifies completely a model of a non-stationary time series $\{X_t\}_{t=1}^{n}$ defined in (\ref{CGINAR}). According to \cite{DavisYau2013}, the MDL for this model is given by
\begin{align*}
\textbf{MDL}(m,\bm{\lambda},\bm{p},\bm{\theta})=&\log(m)+(m+1)\log n+\sum\limits_{j=1}^{m+1}\log(p_j)+\sum\limits_{j=1}^{m+1}\frac{p_j+1}{2}\log(n_j)\\
&-\sum\limits_{j=1}^{m+1}\tilde{L}_{n_j}^{(j)}(\bm{\theta_j};p_j,\bm{x_j}).
\end{align*}
The estimators of the number of change-points, the locations of change-points and the parameters in each of the segments are given by the vector $(\hat{m}_n,\bm{\hat{\lambda}_n},\bm{\hat{p}_n},\bm{\hat{\theta}_n})$, that is
\begin{align}\label{MDL}
(\hat{m}_n,\bm{\hat{\lambda}_n},\bm{\hat{p}_n},\bm{\hat{\theta}_n})=\arg\min\limits_{
m\leq M_0,\bm{\lambda}\in A_{\epsilon_{\lambda}}^m,\hfill\atop
 {\bm{p}\in\mathcal{P},\bm{\theta}\in\bm{\Theta}\hfill}}
 \textbf{MDL}(m,\bm{\lambda},\bm{p},\bm{\theta}),
\end{align}
where $\bm{\hat{\lambda}_n}=(\hat{\lambda}_1,...,\hat{\lambda}_{\hat{m}_n}),~\bm{\hat{p}_n}=(\hat{p}_1,...,\hat{p}_{\hat{m}_n+1}),
~\bm{\hat{\theta}_n}=(\bm{\hat{\theta}_1},...,\bm{\hat{\theta}_j},...,\bm{\hat{\theta}_{\hat{m}_n+1}})$, with
$$\bm{\hat{\theta}_j}=\arg\max\limits_{\bm{\theta_j}\in\Theta_j(\hat{p}_j)}\tilde{L}_{n_j}^{(j)}(\bm{\theta_j};\hat{p}_j,\bm{\hat{x}_j}),$$
and $\bm{\hat{x}_j}=\{x_t\}_{t=[n\hat{\lambda}_{j-1}]+1}^{t=[n\hat{\lambda}_j]}$.

Let $\{x_t\}_{t=1}^n$ be observations from a piecewise stationary process (\ref{CGINAR}) specified by the true value vector $(m^0,\bm{\lambda^0},\bm{p^0},\bm{\theta^0})$, where ${\bm\lambda^0}=(\lambda_1^0,...,\lambda_{m^0}^0)$, $\bm{p^0}=(p_1^0,...,p_{m^0}^0)$, $\bm{\theta^0}=(\bm{\theta}_1^0,...,\bm{\theta}_{m^0}^0)$.
We first give the following Theorem \ref{MDL_mknown_theorem}, which indicates the assumption and the result about the convergence of the change-points locations and the order parameter estimators when the number of change-points is known.
\begin{theorem}\label{MDL_mknown_theorem}
Assume that ${\rm E}|X_{t,j}|^{2+\eta}<\infty$ for some $\eta>0$. Suppose the number of change-points $m^0$ is known and we estimate the locations and the model parameter by
\begin{align}\label{MDL_mknown}
(\bm{\hat{\lambda}_n},\bm{\hat{p}_n},\bm{\hat{\theta}_n})=\arg\min\limits_{
\bm{\lambda}\in A_{\epsilon_{\lambda}}^{m^0},\hfill\atop
 {\bm{p}\in\mathcal{P},\bm{\theta}\in\bm{\Theta}\hfill}}{\bm{{\rm MDL}}}(m^0,\bm{\lambda},\bm{p},\bm{\theta}).
\end{align}
Then, $\bm{\hat{\lambda}_n}\xrightarrow{a.s.}\bm{\lambda^0}$ and for each segment the estimated model does not underestimate the true model.
\end{theorem}
The following two theorems will discuss consistent results under the condition of the number of change-points is unknown.
\begin{theorem}\label{week}
{\rm\bf{(Weak Consistency)}} Assume that ${\rm E}(X_{t,j})^{4+\eta}<\infty$. The estimator $(\hat{m}_n,\bm{\hat{\lambda}_n},$ $\bm{\hat{p}_n}, \bm{\hat{\theta}_n})$ is defined in (\ref{MDL}). Then we have week consistency of ${\bm{{\rm MDL}}}$ model selection, i.e.,
\begin{align*}
\hat{m}_n\xrightarrow{p}m^0,~~~\bm{\hat{\lambda}_n}\xrightarrow{p}\bm{\lambda^0},~~~\bm{\hat{p}_n}\xrightarrow{p}\bm{p^0},~~~{\bm{\hat{\theta}_n}}\xrightarrow{p}{\bm{\theta^0}},
\end{align*}
and for each $\lambda_j^0$, $j=1,...,m^0$, there exists a $\hat{\lambda}_{j'}\in \bm{\hat{\lambda}_n}$, $1\leq j'\leq \hat{m}_n$, such that for any $\kappa>0$,
\begin{align*}
\left|\hat{\lambda}_{j'}-\lambda_j^0\right|=O_p(n^{\kappa-1}).
\end{align*}
\end{theorem}
After strengthening the moment condition, the following theorem shows strong consistency in the MDL model selection process.
\begin{theorem}\label{strong}
{\rm\bf{(Strong Consistency)}} Assume that ${\rm E}(X_{t,j})^{8+\eta}<\infty$. The estimator $(\hat{m}_n, \bm{\hat{\lambda}_n},$ $\bm{\hat{p}_n}, \bm{\hat{\theta}_n})$ is defined in (\ref{MDL}). Then we have strong consistency of ${\bm{{\rm MDL}}}$ model selection, i.e.,
\begin{align*}
\hat{m}_n\xrightarrow{a.s.}m^0,~~~\bm{\hat{\lambda}_n}\xrightarrow{a.s.}\bm{\lambda^0},~~~\bm{\hat{p}_n}\xrightarrow{a.s.}\bm{p^0},~~~{\bm{\hat{\theta}_n}}\xrightarrow{a.s.}{\bm{\theta^0}},
\end{align*}
and for each $\lambda_j^0$, $j=1,...,m^0$, there exists a $\hat{\lambda}_{j'}\in \bm{\hat{\lambda}_n}$, $1\leq j'\leq \hat{m}_n$, such that
\begin{align*}
\left|\hat{\lambda}_{j'}-\lambda_j^0\right|=o(n^{-\frac{1}{2}}),~~~\text{a.s.},
\end{align*}
furthermore, if the order $p_{j'}^0$ of the ${j'}$-th piece true model is specified, then
\begin{align*}
\bm{\hat{\theta}_{j'}}(\hat{\lambda}_{j'-1},\hat{\lambda}_{j'})-\bm{\theta_j^0}=O_p(\sqrt{\frac{\log\log n}{n}}),
\end{align*}
where
\begin{align*}
\bm{\hat{\theta}_{j'}}(\hat{\lambda}_{j'-1},\hat{\lambda}_{j'})&=\arg\max\limits_{\bm{\theta_{j'}}\in\Theta_{j'}(p_{j'}^0)}\tilde{L}_{n_{j'}}^{({j'})}(\bm{\theta_{j'}},\hat{\lambda}_{{j'}-1},\hat{\lambda}_{{j'}};p_{j'}^0,\bm{\hat{x}_{j'}})\\
&=\arg\max\limits_{\bm{\theta_{j'}}\in\Theta_{j'}(p_{j'}^0)}\left(\sum\limits_{t=[\hat{n}_{j'}\hat{\lambda}_{{j'}-1}]+1}^{[\hat{n}_{j'}\hat{\lambda}_{j'}]}\ell_t^{(j')}(\bm{\theta_{j'}};p_{j'}^0, |x_{s,{j'}},s<t)\right).
\end{align*}
\end{theorem}
\begin{remark}
Clearly, from the results obtained in Theorem \ref{week} and Theorem \ref{strong}, the convergence rate of the PQML estimator $\bm{\hat{\theta}}$ is not affected even when the estimated piece may not be fully inside a stationary piece of a time series but involves part of the adjacent stationary pieces.
\end{remark}
\section{Genetic Algorithm}\label{Set4}
Genetic Algorithm (GA), a global search optimization technique, was first proposed by Holland and is based on the principle of natural selection. Holland described this concept in his book, "Adaptation in Natural and Artificial Systems". GA is a population-based search algorithm that applies the survival of the survival of the fittest concept. By applying genetic operators repeatedly to people already existing in the population, new populations are created. The main components of GA are chromosome representation, selection, crossover, mutation, and fitness function computation.

As the era of data arrived, many academics quickly investigated and invented numerous algorithms.
The GA package or function is now available in several software, including the SBRect package for the statistical language R (\cite{Doerr2017}) and the ga function for MATLAB. However, defining special chromosome coding, crossover, and mutation, as well as adding specific optimization processes (such as annealing processes), often helps to improve the accuracy and speed of optimization results, especially for the optimising problems being studied.

In \cite{Davis2006}, a very classical and comprehensive genetic algorithm automatic piecewise autoregressive modelling process (Auto-PARM) was suggested for the optimal MDL criterion. In the article that followed, published in \cite{Aue2014}, the algorithm was still applied for data segmentation and variable selection based on quantile regression. In recent years numerous scholars have addressed their experiences with genetic algorithms. In light of these experiences, we enhanced Auto-PARM in two aspects: modifying the \textbf{crossing operation} and adding the \textbf{annealing process}, which we dubbed automatic piecewise generalized integer autoregressive modeling program with simulated annealing(Auto-PGINARM-SA). Although some of the steps are the same as those of \cite{Davis2006}, we need to repeat them to ensure the integrity of the algorithm for the sake of readability.
\subsection{Chromosome representation}
The representation of a potential solution as a chromosome affects a GA's performance, and for the current issue (minimizing (\ref{MDL})), a chromosome should contain all the information that any $\mathcal{F}\in\mathcal{M}$ could possibly need regarding the change-points, $\tau_j$, and the $j$-th segment GINAR orders, $p_j$. Poisson quasi-maximum likelihood estimators of other model parameters can be uniquely calculated once these quantities have been established. Here we use the same chromosome representation as in \cite{Davis2006}, a chromosome $\delta=(\delta_1,...,\delta_n)$ is of length $n$ with gene values $\delta_t$ defined as
\begin{align*}
\delta_t= \left\{
\begin{array}{ll}
p_j&\text{if $t=\tau_{j-1}$ and the GINAR order for the $j$-th piece is $p_j$,}\\
-1&\text{if no change-point at $t$.}
\end{array} \right.
\end{align*}
In practice, the following constraints are imposed on each $\delta$. First, an upper bound $P_0$ is put on the order $p_j$ of the GINAR process in the algorithm's actual implementation. Given that we require a large $P_0$ to capture complicated sequences and that $P_0$ cannot exceed the number of observations $n$ for simple sequences, there doesn't appear to be a single value for $P_0$. In keeping with past published experience, we decided to set $P_0 = 20$ in our simulation.
At the same time, the following ``minimum span" restriction is imposed to $\delta$ in order to have sufficient observed data for parameter estimation. A piece of $\mathcal{F}$ must have at least $m_p$ observations if the GINAR order of the piece is $p$.
To guarantee that there are enough observations to produce estimators of the GINAR($p$) process parameters, this preset integer $m_p$ was used. Table 1 contains the relevant minimum span $m_p$.
\begin{table}[H]
\vspace{-3mm}
	\caption{Values of $m_p$ for order $p$.}
	\label{mp}
	\vspace{-3mm}                        
	\centering                            
		\begin{tabular}{*{9}{c}}
			\midrule
			$p$&0$-$1&2&3&4&5&6&7$-$10&11$-$20\\\midrule
			$m_p$&10&12&14&16&18&20&25&50\\
			\midrule
		\end{tabular}
\end{table}
\subsection{Initial Population Generation}
For the chromosome $\delta=(\delta_1,...,\delta_n)$, $\delta_t$ will be a change-point with probability $\pi_B=10/n$ and $-1$ with probability $1-\pi_B$. Once $\delta_t$ is declared to be a break, select a value for $p_t$ from $\{0,...,P_0\}$ with equal probabilities and set $\delta_t = p_t$. Then to ensure the minimum span constraint, set the next $m_{p_t}-1$ genes $\delta_i$ (i.e.,$\delta_{t+1},...,\delta_{t+m_{p_t}-1}$) is $-1$. For the next gene $\delta_{t+m_{p_t}}$ we repeat the process of $\delta_t$ until the last gene $\delta_n$. Just to be clear, for the first gene, $\delta_1$, must serve as the beginning point of the first piecewise, thus we choose directly from the uniform distribution with values $\{0,...,P_0\}$ and set $\delta_1 = p_1$.
\subsection{Selection}
Once the first generation of random chromosomes is generated, we select some of the chromosomes with good genes as parents for the following crossover and mutation operations. The probability of the two parents is inversely proportional to their ranking in order of MDL values. In other words, chromosomes with lower MDL values are more likely to be selected. As for the select method, in our simulation, we chose the roulette to select the parents.

After parental selection, crossover and mutation operations are applied to produce offspring, which will form a second generation. Offspring tend to be better than their parents because they are better solutions to optimizing problems.
The probability of carrying out a crossover operation in this implementation is set to $\pi_C$, while the probability of carrying out a mutation operation is set to $1-\pi_C$, where $\pi_C = (n-10)/n$. Next, the crossover and mutation operations used in this paper are introduced respectively.
\subsection{Crossover}
We first introduce the following two crossover operations.

\emph{Uniform crossover}: this operation on two parent strings $\delta_f=(\delta_1^{(1)},...,\delta_n^{(1)})$ and $\delta_m=(\delta_1^{(2)},...,\delta_n^{(2)})$ is defined as follows. The offspring string $\delta_c=(\delta_1^{(3)},...,\delta_n^{(3)})$ is constructed by independently choosing the string entries from the two
parents with equal probabilities, i.e.,
\begin{align*}
\delta_{t}^{(3)}= \left\{
\begin{array}{ll}
\delta_{t}^{(1)}&\text{with probability 1/2;}\\
\delta_{t}^{(2)}&\text{with probability 1/2.}
\end{array} \right.
\end{align*}
Pertaining to the Uniform crossover, for $t = 1$, $\delta_1^{(3)}$ takes on the corresponding value $\delta_1^{(3)}=p_1$ from either the first parent $\delta_1^{(1)}$ or second parent $\delta_1^{(2)}$ with equal probabilities, the next $m_{p_1}-1$ genes are set to $-1$. Then for $t =m_{p_1}+1$, $\delta_{m_{p_1}+1}^{(3)}$ is again taken from the first or second parent with equal probability, If this value is $-1$, then the same gene-inheriting process is repeated for the next gene in line (i.e., $\delta_{m_{p_1}+2}^{(3)}$). If this value is not $-1$, then it is a nonnegative integer $p_j$ denoting the GINAR order of the current piece. In this case the minimum span constraint is imposed (i.e., the next $m_{p_j}-1$ are set to $-1$), and repeat the same genetic process until $\delta_n$.

\emph{One-point crossover}: this operation on two parent strings $\delta_f=(\delta_1^{(1)},...,\delta_n^{(1)})$ and $\delta_m=(\delta_1^{(2)},...,\delta_n^{(2)})$, an index $k$ is chosen uniformly at random from $1,..., n $. The child $\delta_c$ consist of the first $k$ values from $\delta_f$ and the last $n-k$ values from $\delta_m$ , i.e
$$\delta_c=(\delta_1^{(1)},...,\delta_k^{(1)},\delta_{k+1}^{(2)},...,\delta_n^{(2)}).$$
Due to the minimum span constraint, the number of $m_{p_t}-1$ following the change point $\delta_t^{(1)}$ with order $p_t$ should be omitted during the random selection of $k$, which means that $k$ cannot be picked in $\delta_{t+1}^{(1)},...,\delta_{t+m_{p_t}-1}^{(1)}$. 

Currently, the majority of genetic algorithms only employ one sort of crossover, such as uniform crossover, which is adopted by \cite{Davis2006} and \cite{Aue2014}.
However, according to the discussion in \cite{Doerr2017}, uniform crossover significantly raises fitness and one-point crossover supports the building block structure of the underlying problem. Strings generated by mutation and uniform crossover look empirically good, but often have ``a flaw in one or two places".
In contrast, one-point crossover make it possible to integrate the flaw-less parts of two strings to an almost perfect one.
We will apply the two types of crossover operations mentioned above in this study and draw on the knowledge of other researchers who have studied this approach.
Although it is evident from the analysis above that using two different types of crossover is absolutely necessary, we nevertheless believe subjectively that uniform crossover is more secure, hence the uniform crossover is primarily employed in the two techniques. For the two crossover operations, we assume that if the crossing occurs, the probability of a uniform crossing is
$$P(\text{uniform crossover occur})=1-P(\text{one-point crossover occur})=\pi_{cross}=0.7.$$
\subsection{Mutation}
One child is produced from one parent in the case of mutation. This method avoids the issue of too quickly convergent to a suboptimal solution by giving the child chromosome more opportunity to explore the search space. This process starts with the second change-point $t$, and every $\delta_t$ (subject to the minimum span constraint) can take one of the following three possible values: It does three things: (a) with probability $\pi_P$, it takes the parent's corresponding $\delta_t$ value; (b) with probability $\pi_N$, it takes the value $-1$; and (c) with probability $1-\pi_P-\pi_N$, it takes a fresh GINAR order ($p_j$) that was produced at random. In the present implementation, $\pi_P$ and $\pi_N$ are both set to $0.3$.
It should be noted that for $t=1$, only (a) and (c) mutations are carried out, and that $\pi_P=0.5$ and $\pi_N=0.$

\subsection{Simulated Annealing}\label{4.6}
The concept of annealing was first proposed by \cite{Kirkpatrick1983} and developed by Metropolis, and it has proven to be a successful method to address the issue of precocious.
Its main idea is to randomly select a solution in the neighborhood of the optimal solution $\delta$, and then compare them to determine which is best. Taking into account the parameters to be optimized, we introduce the following two annealing operations.

\emph{Annealing of the location of change-points $\tau$.}
This operation on an optimal chromosome of the current generation $\delta=(\delta_1,...,\delta_n)$ performs the following operation to randomly select a solution $\delta_{A_{\tau}}$ in its neighborhood. If $t$ is a change-point with order $p_t$, randomly select an integer $s$ from a range $[a, b]$, makes $\delta_{t+s}=\delta_{t}=p_t$, and then $\delta_{t}=-1$. In the present implementation, $a=-10,b=10$. Of course, if the minimum span constraint cannot be guaranteed after such changes, the operation on $\delta_t$ is abandoned, and the operation on the next change point is repeated until the last change point. And then we're going to get a neighborhood solution $\delta_{A_{\tau}}$.

\emph{Annealing of order $p$.} In this operation, for the $j$-th piece, which contains data ${\bf x_t}=(x_{j_1},...,x_{j_s})$, corresponding to the gene segment in the chromosome $(\delta_{j_1},...,\delta_{j_s})$. Through the Yule-Walker parameter estimation expression in \cite{SilvaSilva2006}, the estimator $(\hat{\alpha}_{j_1},... \hat{\alpha}_{j_{P_0}})$ can be calculated based on $\bf{x_t}$. Then make the following substitution for the change-point location gene, $\delta_{t_1} = \max\limits_u \{\hat{\alpha}_{j_u}>\pi_{A_p}, u=1,2,...,P_0\}$ (here we set $\pi_{A_p}=0.05$). The process is also abandoned if the minimum span constraint cannot be guaranteed following such a change and the subsequent part is repeated until the last piece. And then we're going to get a neighborhood solution $\delta_{A_{p}}$.

In conclusion, perform the following steps for the annealing process.
A new population is produced once the crossover and mutation are finished.
In this population, the chromosome $\delta_1$ with the lowest MDL value was found.
Then, apply the two annealing operations mentioned above to $\delta_1$ to obtain its two neighborhood solutions, $\delta_{A_{\tau}}$ and $\delta_{A_{p}}$. Make $\delta_1'=\arg \min\limits_{\delta \in \{\delta_1,\delta_{A_{\tau}},\delta_{A_p}\}} \text{MDL}(\delta)$, and let $\delta_1=\delta_1'$ be the final step.
\subsection{Elitist step and Island model}
Two more phases, the elitist step and the island model, are carried out to hasten the convergence to an optimal solution.
In the elitist stage, The best chromosome from the present generation will take the place of the worst chromosome from the next generation. This ensures the monotonicity of the search process and conserves the best chromosome in each generation.
The island model is carried out using parallel implementations, and $NI$ (number of islands) GAs are run simultaneously in $NI$ different subpopulations. After every $J$ generations, the best $H$ chromosomes from the ($j-1$)th island will replace the worst $H$ chromosomes from the $j$th island, $j = 2,...,NI$ (for $j=1$ the worst $H$ chromosomes are replaced by the best chromosomes from the $NI$-th island). Here, $NI=40, J=5, H=2$, and a subpopulation size of $40$ are used in the implementation.

\section{Simulation}\label{Set5}
To evaluate the finite-sample performance of the proposed Auto-PGINARM-SA procedure, we conduct extensive simulation studies and split the simulation studies into the following three parts. First, Auto-PGINARM-SA was compared with Auto-PARM, proposed by \cite{Davis2006}. Note that Auto-PARM is for autoregressive process, while the problem we studied is general integer-valued autoregressive process, so we replace the Gaussian quasi likelihood function in Auto-PARM with the PQML function and name it Auto-PGINARM procedure. In the second simulation, MDL was contrasted with the existing method penQLIK (\cite{DiopKengne2021b}). In the last part, Six sets of simulation are utilised to study the consistency conclusion. For convenience, we first introduce the following two INAR($p$) processes that are primarily used in simulations.

The $j$-th piece process $\{X_{t,j}\}$ is said to INAR($p_j$) process based on the binomial thinning operator (BiINAR($p_j$)) if it is determined by the recursion:
\begin{align}\label{BiINAR}
X_{t,j}=\alpha_{1,j}\circ X_{t-1,j}+...+\alpha_{p_j,j}\circ X_{t-p_j,j}+Z_{t,j},
\end{align}
where $``\circ"$ is the binomial thinning operator, which is proposed by \cite{{SV1979}} and defined as: $\alpha\circ X=\sum_{i=1}^{X}\mathrm{B}_{i}(\alpha)$, the counting series $\{\mathrm{B}_i(\alpha)\}$ is an independent and identically distributed (i.i.d.) Bernoulli random sequence with mean $\alpha$, $\{Z_{t,j}\}$ is a sequence of i.i.d. Poisson random variables with mean $\gamma$.

The $j$-th piece process $\{X_{t,j}\}$ is said to INAR($p_j$) process based on the negative binomial thinning operator (NBINAR($p_j$)) if it is determined by the recursion:
\begin{align}\label{NBINAR}
X_{t,j}=\alpha_{1,j}\ast X_{t-1,j}+...+\alpha_{p_j,j}\ast X_{t-p_j,j}+Z_{t,j},
\end{align}
where $``\ast$" is the negative binomial thinning operator of \cite{Ristic2009}, it is defined as $\alpha\ast X=\sum_{i=1}^X \mathrm{W}_i(\alpha)$, the counting series $\{W_i(\alpha)\}$ is a sequence of i.i.d. geometric random variables with parameter $\alpha/(1+\alpha)$, $\{Z_{t,j}\}$ is a sequence of i.i.d. Geometric random variables with parameter $\gamma/(1+\gamma)$.

All simulations are carried out using the MATLAB software. The empirical results displayed in the tables, including the true positive rate (TPR) that is the proportion of informative points are correctly identified, the empirical biases (Bias) and mean square errors (MSE), are computed over $1000$ replications.
\subsection{Efficiency of Simulated Annealing}
The objective of the following scenario MCP-BiINAR is to evaluate the performance of simulated annealing by comparing the optimization efficacy of Auto-PGINARM and Auto-PGINARM-SA.
\begin{align*}
X_{t}=\left\{
\begin{array}{ll}
0.5\circ X_{t-1,1}+Z_{t,1},&1\leq t\leq 400,\\
0.2493\circ X_{t-1,2}+0.254\circ X_{t-2,2}+0.2967\circ X_{t-3,2}+Z_{t,2},&401\leq t\leq 800,\\
0.4\circ X_{t-1,3}+Z_{t,3},&801\leq t\leq 1000,\\
\end{array}\right.
\end{align*}
where $``\circ"$ is defined in (\ref{BiINAR}) and $\{Z_{t,j}\}_{j=1,2,3}$ is a sequence of i.i.d. Poisson random variables with mean $\gamma_j$, and  $(\gamma_1,\gamma_2,\gamma_3)=(0.5,1,0.5).$

Table \ref{compare_anneal} provides a summary of all the findings, including the TPR of the number of change points (${\rm TPR}(m)$) and the order $p$ of each segment (${\rm TPR}(p)$), the Bias and MSE of the change-points locations, the average MDL value ($\overline{\text{MDL}}$), average time(s), annealing frequency, and the relative frequency when the MDL value of Auto-PGINARM-SA was smaller in 1000 replications.
The following results can be seen from the Table \ref{compare_anneal}. Out of 1000 replications, annealing ran 89912 times, demonstrating its effectiveness in Auto-PGINARM-SA. The results of Bias, MSE, and ${\rm TPR}(p)$ demonstrate that Auto-PGINARM-SA is essentially always superior to Auto-PGINARM. Simulated annealing significantly improves the accuracy of MDL optimization, particularly for the estimation of $p_3$ in the third segment.
It is not difficult to see from the average time that the simulated anneal process improves the speed of finding the optimal solution.
In terms of MDL value, the average MDL value calculated by Auto-PGINARM-SA is smaller. Remarkably, Auto-PGINARM-SA was able to find a better solution 982 times out of 1000 replications.
All conclusions indicate that, with the assistance of simulated annealing, Auto-PGINARM-SA could quickly find better solutions and effectively solve the precocious problem.
\begin{table}[H]
\vspace{-3mm}
\caption{Summary of the performance of simulated annealing by comparing the results of Auto-PGINARM and Auto-PGINARM-SA under the scenario MCP-BiINAR.}\label{compare_anneal}
\vspace{-3mm}
\begin{adjustwidth}{-1cm}{-1cm}
{\tabcolsep0.05in
\begin{tabular}{crrcccrrccc}
\midrule
&\multicolumn{5}{c}{Auto-PGINARM}&\multicolumn{5}{c}{Auto-PGINARM-SA}\\\cmidrule(lr){2-6}\cmidrule(lr){7-11}
&\multicolumn{1}{c}{Bias}&\multicolumn{1}{c}{MSE}&${\rm TPR}(p)$&${\rm TPR}(m)$&$\overline{\text{MDL}}$&\multicolumn{1}{c}{Bias}&\multicolumn{1}{c}{MSE}&${\rm TPR}(p)$&${\rm TPR}(m)$&$\overline{\text{MDL}}$\\
$p_1$&0.0000 &0.0000 &\bf{1}&0.99&$-$713.5010 &0.0000 &0.0000 &\bf{1}&0.989&$-$716.6701 \\
$p_2$&$-$0.0020 &0.0040 &\bf{0.996}&&&0.0000 &0.0020 &\bf{0.998}&&\\
$p_3$&0.1960 &0.2404 &\bf{0.798}&&&0.0060 &0.0320 &\bf{0.974}&&\\
$\lambda_1$&$-$0.0013 &0.0001 &&&&$-$0.0003 &0.0001 &&&\\
$\lambda_2$&$-$0.0007 &0.0000 &&&&$-$0.0011 &0.0000 &&&\\\cmidrule(lr){2-6}\cmidrule(lr){7-11}
&\multicolumn{4}{l}{Average time(s): 2595.251}&&\multicolumn{5}{l}{Average time(s): 2171.780}\\
\midrule
\multicolumn{10}{l}{Annealing frequency}&\bf{89812}\\
\multicolumn{10}{l}{the relative frequency of MDL(Auto-PGINARM-SA)$<$MDL(Auto-PGINARM)}&\bf{0.982} \\
\midrule
\end{tabular}
}
\end{adjustwidth}
\end{table}
\subsection{Comparison between the MDL and penQLIK}
Next, under the following scenario MCP-NBINAR, we make a comparison of MDL with the penalized contrast QLIK (penQLIK) (\cite{DiopKengne2021b}) which  is carried out a data-driven method, based on the slope heuristic procedure (see \cite{Baudry2012}) to calibrate the penalty term. The dynamic programming algorithm is used to minimize the penQLIK criteria.
\begin{align*}
\footnotesize
X_{t}=\left\{
\begin{array}{ll}
0.5\ast X_{t-1,1}+Z_{t,1},&1\leq t\leq 300,\\
0.4524\ast X_{t-1,2}+0.1818\ast X_{t-2,2}+0.1658\ast X_{t-3,2}+Z_{t,2},&301\leq t\leq 500,\\
0.4\ast X_{t-1,3}+Z_{t,3},&501\leq t\leq 800,\\
0.0252\ast X_{t-1,4}+0.1502\ast X_{t-2,4}+0.4692\ast X_{t-3,4}+0.1554\ast X_{t-4,4}+Z_{t,4}&801\leq t\leq 1000,
\end{array}\right.
\end{align*}
where $``\ast"$ is the negative binomial thinning operator, defined in (\ref{NBINAR}), and $\{Z_{t,j}\}_{j=1,2,3,4}$ is a sequence of i.i.d. Geometric random variables with mean $\gamma_j$, respectively, and $(\gamma_1,\gamma_2,\gamma_3,\gamma_4)=(0.5,1,0.5,2).$

\begin{figure}[H]
\begin{adjustwidth}{-2cm}{-1cm}
\includegraphics[width=8.5in,height=2.2in]{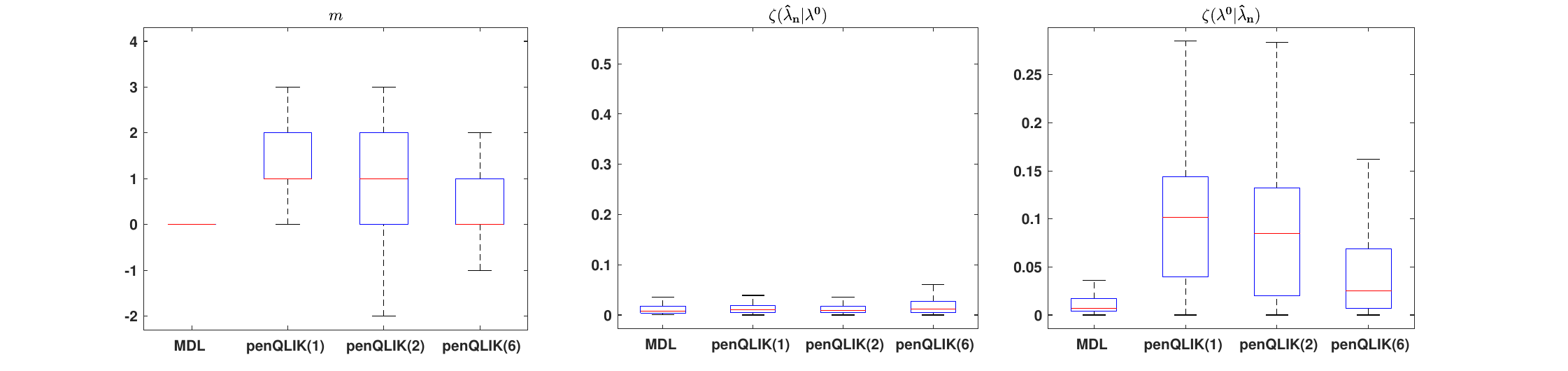}
\end{adjustwidth}
\vspace{-3mm}
\caption{Boxplot of 1000 MDL, penQLIK(1), penQLIK(2) and penQLIK(6) simulations result for the Bias of the number of change-points estimator (left), the over-segmentation error(middle),  the under-segmentation error (right) under the scenario MCP-NBINAR.}
\label{box_pen}
\end{figure}

Considering that the order $p$ is not taken into account in the information criterion of penQLIK, and a specific $p$ value cannot be set in the scenario MCP-NBINAR, we respectively simulated the effect of penQLIK in the three cases of $p=1,2,6$. They are denoted as penQLIK(1), penQLIK(2), penQLIK(6).
In this section we not only calculate the ${\rm TPR}$ of the number of change-points, the Bias and MSE of the location of change-points, but also calculate the distance between the estimated set $\bm{\hat{\lambda}_n}$ and the true change-point set $\bm{\lambda^0}$ (\cite{Boysen2009}),
\begin{align*}
\zeta(\bm{\hat{\lambda}_n|\lambda^0})=\sup\limits_{b\in\bm{\lambda^0}}\inf\limits_{a\in\bm{\hat{\lambda}_n}}|a-b|,
~~~\zeta(\bm{\lambda^0|\hat{\lambda}_n})=\sup\limits_{b\in\bm{\hat{\lambda}_n}}\inf\limits_{a\in\bm{\lambda^0}}|a-b|.
\end{align*}
which quantify the over-segmentation error and the under-segmentation error, respectively. A desirable estimator should be able to balance both quantities. All results are summarized in Table \ref{compare_pen}. To make things more intuitive, we also present the boxplot (Figure \ref{box_pen}), which is obtained from the bias of $m$, the average values of $\zeta(\bm{\hat{\lambda}_n|\lambda^0})$, $\zeta(\bm{\lambda^0|\hat{\lambda}_n})$ simulation results for  the scenario MCP-NBINAR.

Clearly, different values of $p$ have great influence on the performance of penQLIK. As can be seen from the results in the Table \ref{compare_pen} and boxplot, although the effect of penQLIK is significantly improved when $p$ value is relatively large, there is still a big gap between the effect of penQLIK and MDL. More intuitively, from boxplot, the $\zeta(\bm{\hat{\lambda}_n|\lambda^0})$ and $\zeta(\bm{\lambda^0|\hat{\lambda}_n})$ of penQLIK(1) and penQLIK (2)  are out of balance, that is, serious under-segmentation error will occur in these two information criterion. Although penQLIK(6) obviously reduces the occurrence of the under-segmentation error, the performance is not as good as that of MDL.
\begin{table}[H]
\caption{Comparison results of MDL and penQLIK under the scenario MCP-NBINAR.}\label{compare_pen}
\vspace{-3mm}
\begin{adjustwidth}{-0.5cm}{-1cm}
{\tabcolsep0.015in
\begin{tabular}{lccccrrrrrrr}
\midrule
&&&&&\multicolumn{3}{c}{$\bm{\hat{\lambda}_n}$}&\multicolumn{4}{c}{$\bm{\hat{p}_n}$}\\
\cmidrule(lr){6-8}\cmidrule(lr){9-12}
Criteria&${\rm TPR}(m)$&$\zeta(\bm{\hat{\lambda}_n|\lambda^0})$&$\zeta(\bm{\lambda^0|\hat{\lambda}_n})$&&\multicolumn{1}{c}{$\hat{\lambda}_1$}&\multicolumn{1}{c}{$\hat{\lambda}_2$}&\multicolumn{1}{c}{$\hat{\lambda}_3$}&\multicolumn{1}{c}{$\hat{p}_1$}&\multicolumn{1}{c}{$\hat{p}_2$}&\multicolumn{1}{c}{$\hat{p}_3$}&\multicolumn{1}{c}{$\hat{p}_4$}\\
\cmidrule{1-1}\cmidrule(lr){2-2}\cmidrule(lr){3-4}\cmidrule(lr){5-12}
MDL&0.949&0.0210 &0.0132 &Bias&0.0020 &$-$0.0020 &0.0004 &0.0200 &$-$0.5290 &0.0948 &$-$0.5100 \\
&&&&MSE&0.0002 &0.0001 &0.0000 &0.0242 &0.6512 &0.0991 &0.5227 \\
penQLIK(1)&0.206&0.0185 &0.0946 &Bias&0.0089 &0.0012 &0.0062 &&&&\\
&&&&MSE&0.0015 &0.0014 &0.0002 &&&&\\
penQLIK(2)&0.296&0.0153 &0.0816 &Bias&0.0058 &0.0022 &0.0046 &&&&\\
&&&&MSE&0.0006 &0.0010 &0.0001 &&&&\\
penQLIK(6)&0.613&0.0285 &0.0482 &Bias&0.0013 &0.0031 &0.0006 &&&&\\
&&&&MSE&0.0010 &0.0010 &0.0001 &&&&\\
\midrule
\end{tabular}
}
\end{adjustwidth}
\end{table}

\subsection{Consistency Analysis}
In this section, we illustrate the ``consistency"-like property of Auto-PGINARM-SA via studying the finite-sample performance on a range of simulated data. we choose BiINAR and NBINAR process as GINAR processes within the MCP-GINAR models, and consider $m^0=1,2,3$ cases. The sample sizes considered are $n =500,1000$ and $1500$.
For the generation of the $j$-th segment coefficient $\bm{\alpha_j}=(\alpha_{1,j},... , \alpha_{p_j, j}) $, we use the following steps.
let $\tilde{\alpha}_j=\sum_{i=1}^{p_j}\alpha_{i,j}$, considering the condition $0<\tilde{\alpha}_j<1$, when we choose a specific $\tilde{\alpha}_j$, we use the "drchrnd" function in MATLAB to generate a list of random numbers $(\tilde{\alpha}_{1,j},... , \tilde{\alpha}_{p_j, j})$, then use $\alpha_{i,j}^0=\tilde{\alpha}_j*\tilde{\alpha}_{i,j},i=1... ,p_j$ as the $j$-th segment's true coefficient. All parameter settings and model definitions are summarized in Table \ref{para_set}.
\begin{table}[H]
\scriptsize
\caption{Parameters setting of different scenarios.}\label{para_set}
\vspace{-3mm}
\centering
\renewcommand\arraystretch{1}
{\tabcolsep0.06in
\begin{tabular}{*{14}{c}}
\midrule
\textbf{Scenarios}&\multicolumn{13}{c}{Parameter Setting}\\\midrule
&\multicolumn{13}{l}{the $j$-th segment: $X_{t,j}=\alpha_{1,j}\circ X_{t-1,j}+...+\alpha_{p_j,j}\circ X_{t-p_j,j}+Z_{t,j}$,}\\
&\multicolumn{13}{l}{where $Z_{t,j}$ is Poisson random variables with parameter $\gamma_j$.}\\\midrule
MCP-BiINAR$_1$&\multicolumn{13}{l}{$m^0=1, \bm{\lambda^0}=0.4n, \bm{p^0}=(1,3), \bm{\theta^0}=(\bm{\theta_1^0},\bm{\theta_2^0})$}\\\cmidrule{2-14}
Segment&\multicolumn{2}{c}{I: BiINAR(1)}&\multicolumn{4}{c}{II: BiINAR(3)}&&&&&&&\\\cmidrule(r){2-3}\cmidrule(lr){4-7}
&$\alpha_{1,1}^0$&$\gamma_1^0$&$\alpha_{1,2}^0$&$\alpha_{2,2}^0$&$\alpha_{3,2}^0$&$\gamma_2^0$&&&&&&&\\
&0.5&0.5&0.4877 &0.0200 &0.2923 &1&&&&&&&\\\cmidrule{2-14}
MCP-BiINAR$_2$&\multicolumn{13}{l}{$m^0=2,\bm{\lambda^0}=(0.4n,0.8n),\bm{p^0}=(1,3,1),\bm{\theta^0}=(\bm{\theta_1^0},\bm{\theta_2^0},\bm{\theta_3^0})$}\\\cmidrule{2-14}
Segment&\multicolumn{2}{c}{I: BiINAR(1)}&\multicolumn{4}{c}{II: BiINAR(3)}&\multicolumn{2}{c}{III: BiINAR(1)}&&&&&\\\cmidrule(r){2-3}\cmidrule(lr){4-7}\cmidrule(lr){8-9}
&$\alpha_{1,1}^0$&$\gamma_1^0$&$\alpha_{1,2}^0$&$\alpha_{2,2}^0$&$\alpha_{3,2}^0$&$\gamma_2^0$&$\alpha_{1,3}^0$&$\gamma_3^0$&&&&&\\
&0.5&0.5&0.1264 &0.1052 &0.5684 &1&0.4&2&&&&&\\\cmidrule{2-14}
MCP-BiINAR$_3$&\multicolumn{13}{l}{$m^0=3,\bm{\lambda^0}=(0.3n,0.5n,0.8n),\bm{p^0}=(1,3,1,4),\bm{\theta^0}=(\bm{\theta_1^0},\bm{\theta_2^0},\bm{\theta_3^0},\bm{\theta_4^0})$}\\\cmidrule{2-14}
Segment&\multicolumn{2}{c}{I: BiINAR(1)}&\multicolumn{4}{c}{II: BiINAR(3)}&\multicolumn{2}{c}{III: BiINAR(1)}&\multicolumn{5}{c}{IV: BiINAR(4)}\\\cmidrule(r){2-3}\cmidrule(lr){4-7}\cmidrule(lr){8-9}\cmidrule(lr){10-14}
&$\alpha_{1,1}^0$&$\gamma_1^0$&$\alpha_{1,2}^0$&$\alpha_{2,2}^0$&$\alpha_{3,2}^0$&$\gamma_2^0$&$\alpha_{1,3}^0$&$\gamma_3^0$&$\alpha_{1,4}^0$&$\alpha_{2,4}^0$&$\alpha_{3,4}^0$&$\alpha_{4,4}^0$&$\gamma_4^0$\\
&0.5&0.05&0.1524 &0.2818 &0.3658 &1&0.2&2&0.0252 &0.0502 &0.5692 &0.2054 &3\\\midrule\midrule
\textbf{Scenarios}&\multicolumn{13}{c}{Parameter Setting}\\\midrule
&\multicolumn{13}{l}{the $j$-th segment: $X_{t,j}=\alpha_{1,j}\ast X_{t-1,j}+...+\alpha_{p_j,j}\ast X_{t-p_j,j}+Z_{t,j}$,}\\
&\multicolumn{13}{l}{where $Z_{t,j}$ is Geometric random variables with parameter $\gamma_j/(1+\gamma_j)$.}\\\midrule
MCP-NBINAR$_1$&\multicolumn{13}{l}{$m^0=1,\bm{\lambda^0}=0.4n,\bm{p^0}=(1,3), \bm{\theta^0}=(\bm{\theta_1^0},\bm{\theta_2^0})$}\\\cmidrule{2-14}
Segment&\multicolumn{2}{c}{I: NBINAR(1)}&\multicolumn{4}{c}{II: NBINAR(3)}&&&&&&&\\\cmidrule(r){2-3}\cmidrule(lr){4-7}
&$\alpha_{1,1}^0$&$\gamma_1^0$&$\alpha_{1,2}^0$&$\alpha_{2,2}^0$&$\alpha_{3,2}^0$&$\gamma_2^0$&&&&&&&\\
&0.5&0.5&0.4877 &0.0200 &0.2923 &1&&&&&&&\\\cmidrule{2-14}
MCP-NBINAR$_2$&\multicolumn{13}{l}{$m^0=2,\bm{\lambda^0}=(0.4n,0.8n),\bm{p^0}=(1,3,1),\bm{\theta^0}=(\bm{\theta_1^0},\bm{\theta_2^0},\bm{\theta_3^0})$}\\\cmidrule{2-14}
Segment&\multicolumn{2}{c}{I: NBINAR(1)}&\multicolumn{4}{c}{II: NBINAR(3)}&\multicolumn{2}{c}{III: NBINAR(1)}&&&&&\\\cmidrule(r){2-3}\cmidrule(lr){4-7}\cmidrule(lr){8-9}
&$\alpha_{1,1}^0$&$\gamma_1^0$&$\alpha_{1,2}^0$&$\alpha_{2,2}^0$&$\alpha_{3,2}^0$&$\gamma_2^0$&$\alpha_{1,3}^0$&$\gamma_3^0$&&&&&\\
&0.5&0.5&0.1264 &0.1052 &0.5684 &1&0.4&2&&&&&\\\cmidrule{2-14}
MCP-NBINAR$_3$&\multicolumn{13}{l}{$m^0=3,\bm{\lambda^0}=(0.3n,0.5n,0.8n),\bm{p^0}=(1,3,1,4),\bm{\theta^0}=(\bm{\theta_1^0},\bm{\theta_2^0},\bm{\theta_3^0},\bm{\theta_4^0})$}\\\cmidrule{2-14}
Segment&\multicolumn{2}{c}{I: NBINAR(1)}&\multicolumn{4}{c}{II: NBINAR(3)}&\multicolumn{2}{c}{III: NBINAR(1)}&\multicolumn{5}{c}{IV: NBINAR(4)}\\\cmidrule(r){2-3}\cmidrule(lr){4-7}\cmidrule(lr){8-9}\cmidrule(lr){10-14}
&$\alpha_{1,1}^0$&$\gamma_1^0$&$\alpha_{1,2}^0$&$\alpha_{2,2}^0$&$\alpha_{3,2}^0$&$\gamma_2^0$&$\alpha_{1,3}^0$&$\gamma_3^0$&$\alpha_{1,4}^0$&$\alpha_{2,4}^0$&$\alpha_{3,4}^0$&$\alpha_{4,4}^0$&$\gamma_4^0$\\
&0.5&0.5&0.4524 &0.1818 &0.1658 &1&0.4&0.5&0.0252 &0.1502 &0.4692 &0.1554 &2\\
\midrule
\end{tabular}
}
\end{table}

\begin{figure}[H]
\begin{adjustwidth}{-2.1cm}{-1cm}
\includegraphics[width=7.6in,height=5.8in]{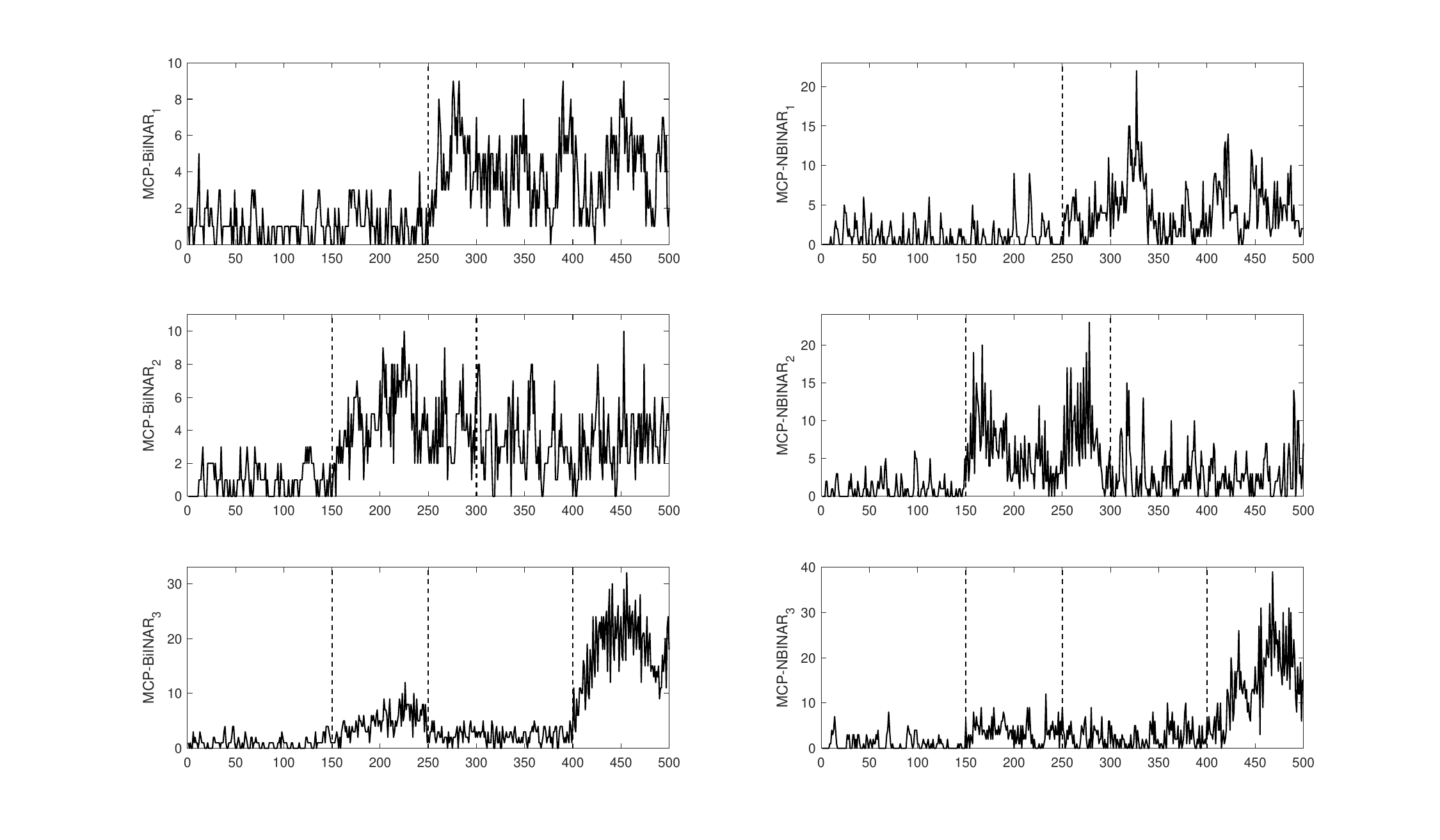}
\end{adjustwidth}
\vspace{-3mm}
\caption{Sample paths plots of the scenarios MCP-BiINAR$_1$ to MCP-BiINAR$_3$. \protect\\ The dotted lines represent the location of the change-points.}
\label{sample-plot}
\end{figure}

We implement sample path of six processes with $n=500$ in Figure \ref{sample-plot}, where all change-points are represented by dotted lines in the figure. We find that under the same parameter values, it is easier to distinguish the change-points regions in MCP-BiINAR processes.
To show a ``consistency"-like property of Auto-PGINARM-SA, we split the simulation results into the following three parts:

(i) Table \ref{para_m} and Table \ref{para_p} indicates the frequencies of the true number of change-points and the MCP-GINAR order $\bm p$. As expected, the simulation results are quite satisfactory.  Although some under-segmentation errors occur when $m^0=3$ and $n=500$, with increasing sample size, the fraction of such ``error" decreases. In general, for $m$ and $\bm p$ it is consistent with our conclusion in Section \ref{Set3}.

(ii) Table \ref{para_lambda_right} indicates the Bias and MSE of the location of change-points  when the number of change-points $m$ is known.
In addition, to evaluate the location accuracy of estimated change-points, we also report the distance from the estimated set $\bm{\hat{\lambda}_n}$ and the true change-point set $\bm{\lambda^0}$ under $m$ is unknown,
$$d(\bm{\hat{\lambda}_n},\bm{\lambda^0})=\frac{1}{|\bm{\lambda^0}|}\sum\limits_{\lambda_k^0\in\bm{\lambda^0}}\min\limits_{\hat{\lambda}_j\in\bm{\hat{\lambda}_n}}|\hat{\lambda}_j-\lambda_k^0| ~~~\text{(\cite{Chen2021})}.$$
Table \ref{para_lambda_right} shows that when $m$ is known, the Bias and MSE of $\bm{\hat{\lambda}_n}$ are both quite minimal, which is a fantastic and satisfying effect and also supports the conclusion of our Theorem \ref{MDL_mknown_theorem}.
On the other hand, it can be shown from $d(\bm{\hat{\lambda}_n},\bm{\lambda^0})$ that the distance considerably reduces with an increase in sample size, which is entirely compatible with the conclusion of consistency in Theorems \ref{week} and \ref{strong}.

\begin{figure}[H]
\begin{adjustwidth}{-2cm}{-1cm}
\includegraphics[width=8.5in,height=8in]{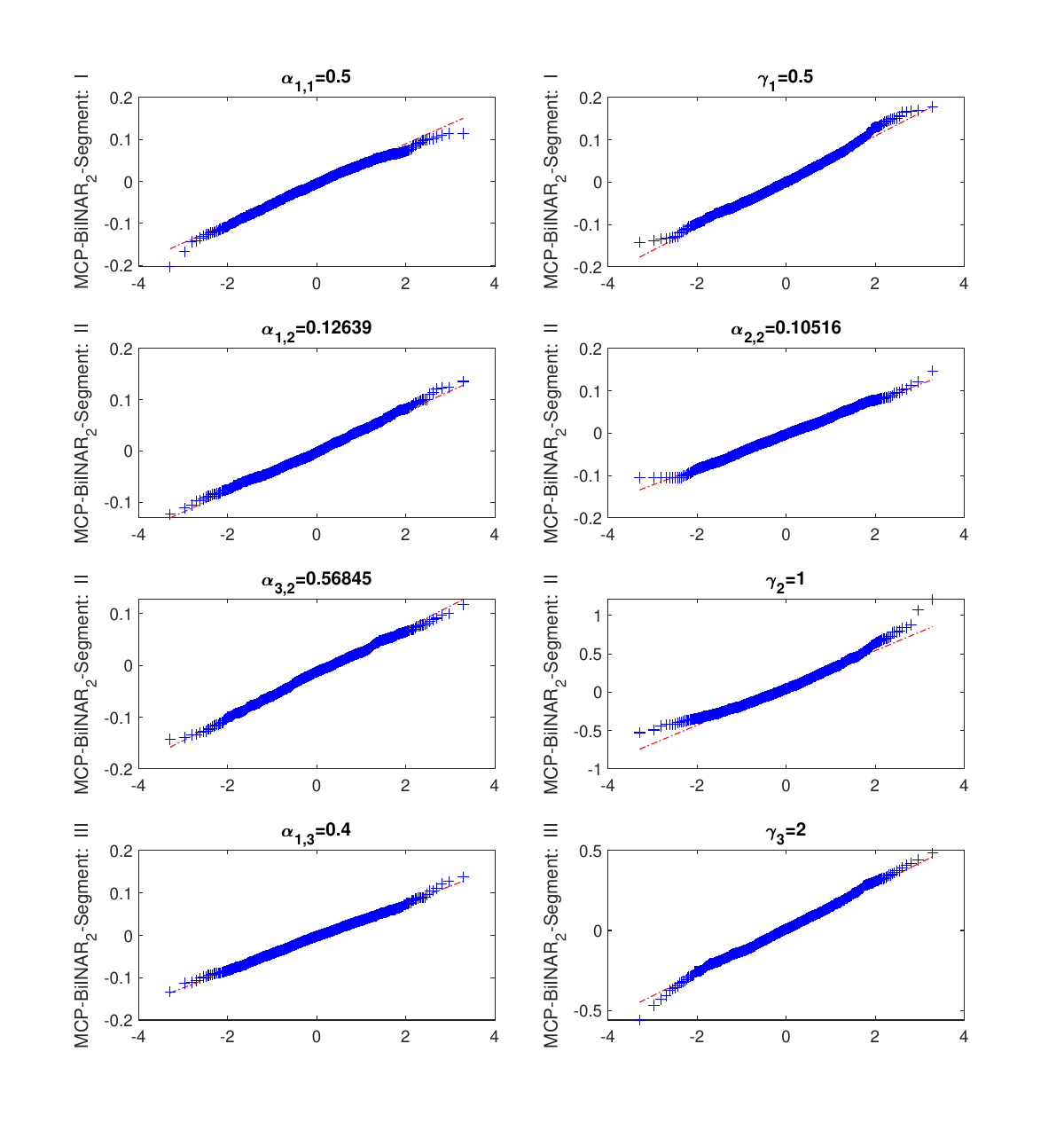}
\end{adjustwidth}
\vspace{-5mm}
\caption{QQ plots of the parameter $\bm{\theta}$ estimator for the scenario MCP-NBINAR$_2$ with sample size $n = 1500$.}
\label{qq-plot}
\end{figure}

(iii) Table \ref{para_alpha} indicates the Bias and MSE of the estimator $\bm{\hat{\theta}}$. Considering that the distance between $\bm{\hat{\theta}}$ and $\bm{\theta^0}$ cannot be effectively defined when $m$ is incorrectly estimated, we only look at the Bias and MSE of $\bm{\hat{\theta}}$ when $m$ is correctly estimated. Tables \ref{para_alpha} shows that the Bias and MSE of $\bm{\hat{\theta}}$ decrease as the sample size $n$ increases. To gain more insight, we also present the QQ plots of the estimator $\bm{\hat{\theta}}$ for MCP-NBINAR$_2$ with sample size $n = 1500$ in Figure \ref{qq-plot}. The QQ plots show that the estimator $\bm{\hat{\theta}}$ is asymptotically normal, as predicted. Similar results are obtained for the remaining models, the figures are omitted to save space.

In conclusion, under all the scenarios, it is evident that the Auto-PGINARM-SA can solve the model selection problem of MCP-GINAR process effectively and perform more stable compared to other model selection method. This is consistent with our theoretical analysis in Section \ref{Set3}. Furthermore, it is worth noting that Auto-PGINARM-SA has superior efficiency for the case with MCP-BiINAR, since the PQML function is very close to its true likelihood function.

\begin{table}[H]
\caption{Frequencies of the number of change-points estimated by the Auto-PGINARM-SA procedure for the scenarios MCP-BiINAR$_1$ to MCP-BiINAR$_3$.}\label{para_m}
\vspace{-3mm}
\begin{adjustwidth}{-1cm}{-1cm}
{\tabcolsep0.05in
\begin{tabular}{*{10}c}
\midrule
&&\multicolumn{3}{c}{Frequencies}&&&\multicolumn{3}{c}{Frequencies}\\\cmidrule(lr){3-5}\cmidrule(lr){8-10}
Scenarios&$n$&$\hat{m}<m^0$&$\hat{m}=m^0$&$\hat{m}>m^0$&Scenarios&$n$&$\hat{m}<m^0$&$\hat{m}=m^0$&$\hat{m}>m^0$\\
\midrule
MCP-BiINAR$_1$&500&0&\bf{1}&0&MCP-NBINAR$_1$&500&0&\bf{0.954}&0.046\\
($m^0=1$)&1000&0&\bf{1}&0&($m^0=1$)&1000&0&\bf{0.982}&0.018\\
&1500&0&\bf{1}&0&&1500&0&\bf{0.989}&0.011\\\cmidrule(lr){2-10}
MCP-BiINAR$_2$&500&0.099&\bf{0.901}&0&MCP-NBINAR$_2$&500&0.011&\bf{0.921}&0.068\\
($m^0=2$)&1000&0&\bf{1}&0&($m^0=2$)&1000&0&\bf{0.931}&0.069\\
&1500&0&\bf{1}&0&&1500&0&\bf{0.933}&0.067\\\cmidrule(lr){2-10}
MCP-BiINAR$_3$&500&0.456&\bf{0.544}&0&MCP-NBINAR$_3$&500&0.44&\bf{0.542}&0.018\\
($m^0=3$)&1000&0.018&\bf{0.982}&0&($m^0=3$)&1000&0.032&\bf{0.949}&0.019\\
&1500&0&\bf{1}&0&&1500&0.002&\bf{0.974}&0.024\\
\midrule
\end{tabular}
}
\end{adjustwidth}
\end{table}

\begin{table}[H]
\scriptsize
\vspace{-3mm}
\caption{Frequencies of the GINAR order estimated by the Auto-PGINARM-SA procedure for the scenarios MCP-BiINAR$_1$ to MCP-BiINAR$_3$.}\label{para_p}
\vspace{-3mm}
\centering
\renewcommand\arraystretch{1.2}
{\tabcolsep0.05in
\begin{tabular}{*{14}c}
\midrule
&&\multicolumn{12}{c}{Frequencies}\\\cmidrule(lr){3-14}
&&\multicolumn{3}{c}{$p_1^0=1$}&\multicolumn{3}{c}{$p_2^0=3$}&\multicolumn{3}{c}{$p_3^0=1$}&\multicolumn{3}{c}{$p_4^0=4$}\\
\cmidrule(lr){3-5}\cmidrule(lr){6-8}\cmidrule(lr){9-11}\cmidrule(lr){12-14}
Scenarios&$n$&\multicolumn{1}{c}{$\hat{p}_1<p_1^0$}&\multicolumn{1}{c}{$\hat{p}_1=p_1^0$}&\multicolumn{1}{c}{$\hat{p}_1>p_1^0$}&\multicolumn{1}{c}{$\hat{p}_2<p_2^0$}&\multicolumn{1}{c}{$\hat{p}_2=p_2^0$}&\multicolumn{1}{c}{$\hat{p}_2>p_2^0$}&\multicolumn{1}{c}{$\hat{p}_3<p_3^0$}&\multicolumn{1}{c}{$\hat{p}_3=p_3^0$}&\multicolumn{1}{c}{$\hat{p}_3>p_3^0$}&\multicolumn{1}{c}{$\hat{p}_4<p_4^0$}&\multicolumn{1}{c}{$\hat{p}_4=p_4^0$}&\multicolumn{1}{c}{$\hat{p}_4>p_4^0$}\\
\cmidrule{1-14}
MCP-BiINAR$_1$&500&0&\bf{1}&0&0.232&\bf{0.766}&0.002&&&&&&\\
($m^0=1$)&1000&0&\bf{0.998}&0.002&0.013&\bf{0.987}&0&&&&&&\\
&1500&0&\bf{1}&0&0.001&\bf{0.994}&0.005&&&&&&\\
\cmidrule(lr){2-14}
MCP-BiINAR$_2$&500&0&\bf{0.99}&0.01&0.03&\bf{0.969}&0.001&0.099&\bf{0.898}&0.003&&&\\
($m^0=2$)&1000&0&\bf{0.999}&0.001&0&\bf{0.999}&0.001&0&\bf{0.997}&0.003&&&\\
&1500&0&\bf{0.999}&0.001&0&\bf{1}&0&0&\bf{0.995}&0.005&&&\\
\cmidrule(lr){2-14}
MCP-BiINAR$_3$&500&0&\bf{0.944}&0.056&0.321&\bf{0.654}&0.025&0.1&\bf{0.534}&0.366&0.958&\bf{0.042}&0\\
($m^0=3$)&1000&0&\bf{1}&0&0.019&\bf{0.978}&0.003&0&\bf{0.963}&0.037&0.707&\bf{0.293}&0\\
&1500&0&\bf{0.999}&0.001&0.001&\bf{0.998}&0.001&0&\bf{0.978}&0.022&0.464&\bf{0.536}&0\\
\midrule
MCP-NBINAR$_1$&500&0&\bf{0.99}&0.01&0.062&\bf{0.918}&0.02&&&&&&\\
($m^0=1$)&1000&0&\bf{0.99}&0.01&0.011&\bf{0.975}&0.014&&&&&&\\
&1500&0&\bf{0.991}&0.009&0.006&\bf{0.977}&0.017&&&&&&\\
\cmidrule(lr){2-14}
MCP-NBINAR$_2$&500&0&\bf{0.982}&0.018&0.012&\bf{0.973}&0.015&0.011&\bf{0.95}&0.039&&&\\
($m^0=2$)&1000&0&\bf{0.99}&0.01&0.002&\bf{0.983}&0.015&0&\bf{0.953}&0.047&&&\\
&1500&0&\bf{0.982}&0.018&0.001&\bf{0.987}&0.012&0&\bf{0.945}&0.055&&&\\
\cmidrule(lr){2-14}
MCP-NBINAR$_3$&500&0&\bf{0.677}&0.323&0.534&\bf{0.395}&0.071&0.286&\bf{0.477}&0.237&0.892&\bf{0.107}&0.001\\
($m^0=3$)&1000&0&\bf{0.956}&0.044&0.487&\bf{0.501}&0.012&0.009&\bf{0.875}&0.116&0.54&\bf{0.454}&0.006\\
&1500&0&\bf{0.989}&0.011&0.314&\bf{0.672}&0.014&0.001&\bf{0.908}&0.091&0.354&\bf{0.637}&0.009\\
\midrule
\end{tabular}
}
\end{table}

\begin{table}[H]
\scriptsize
\vspace{-3mm}
\caption{Summary of the change-points location estimated by the Auto-PGINARM-SA procedure for the scenarios MCP-BiINAR$_1$ to MCP-BiINAR$_3$.}\label{para_lambda_right}
\vspace{-3mm}
\centering
\renewcommand\arraystretch{1.2}
{\tabcolsep0.03in
\begin{tabular}{cccrrrrccccrrrc}
\midrule
Scenarios&$n$&&\multicolumn{1}{c}{$\hat{\lambda}_1$}&\multicolumn{1}{c}{$\hat{\lambda}_2$}&\multicolumn{1}{c}{$\hat{\lambda}_3$}&$d(\bm{\hat{\lambda}_n},\bm{\lambda^0})$&&Scenarios&$n$&&\multicolumn{1}{c}{$\hat{\lambda}_1$}&\multicolumn{1}{c}{$\hat{\lambda}_2$}&\multicolumn{1}{c}{$\hat{\lambda}_3$}&$d(\bm{\hat{\lambda}_n},\bm{\lambda^0})$\\
\cmidrule(lr){4-7}\cmidrule(lr){12-15}
MCP-BiINAR$_1$&500&Bias&0.0003&&&0.0377&&MCP-NBINAR$_1$&500&Bias&0.0018&&&0.0655\\
($m^0=1$)&&MSE&0.0002&&&&&($m^0=1$)&&MSE&0.0009&&&\\
&1000&Bias&$-$0.0006&&&0.0094&&&1000&Bias&0.0007&&&0.0140\\
&&MSE&0.0000&&&&&&&MSE&0.0001&&&\\
&1500&Bias&$-$0.0004&&&0.0043&&&1500&Bias&0.0008&&&0.0065\\
&&MSE&0.0000&&&&&&&MSE&0.0001&&&\\
\cmidrule(lr){1-7}\cmidrule(lr){9-15}
MCP-BiINAR$_2$&500&Bias&$-$0.0001&$-$0.0030&&0.2974&&MCP-NBINAR$_2$&500&Bias&0.0028&0.0007&&0.2900\\
($m^0=2$)&&MSE&0.0002&0.0006&&&&($m^0=2$)&&MSE&0.0003&0.0016&&\\
&1000&Bias&$-$0.0005&$-$0.0013&&0.1500&&&1000&Bias&0.0013&$-$0.0007&&0.1474\\
&&MSE&0.0000&0.0002&&&&&&MSE&0.0001&0.0003&&\\
&1500&Bias&$-$0.0004&$-$0.0005&&0.0997&&&1500&Bias&0.0007&$-$0.0002&&0.0983\\
&&MSE&0.0000&0.0001&&&&&&MSE&0.0000&0.0001&&\\
\cmidrule(lr){1-7}\cmidrule(lr){9-15}
MCP-BiINAR$_3$&500&Bias&0.0013&$-$0.0057&$-$0.0017&0.6056&&MCP-NBINAR$_3$&500&Bias&0.0068&$-$0.0042&0.0013&0.5851\\
($m^0=3$)&&MSE&0.0002&0.0006&0.0001&&&($m^0=3$)&&MSE&0.0009&0.0008&0.0001&\\
&1000&Bias&0.0000&$-$0.0021&$-$0.0007&0.3037&&&1000&Bias&0.0020&$-$0.0020&0.0004&0.3005\\
&&MSE&0.0000&0.0002&0.0000&&&&&MSE&0.0002&0.0001&0.0000&\\
&1500&Bias&0.0000&$-$0.0015&$-$0.0004&0.2023&&&1500&Bias&0.0007&$-$0.0023&$-$0.0012&0.2008\\
&&MSE&0.0000&0.0001&0.0000&&&&&MSE&0.0001&0.0002&0.0004&\\
\midrule
\end{tabular}
}
\end{table}

\begin{table}[H]
\scriptsize
\vspace{-3mm}
\caption{ Summary of parameter estimates obtained by Auto-PGINARM-SA procedure for the scenarios MCP-BiINAR$_1$ to MCP-BiINAR$_3$.}\label{para_alpha}
\vspace{-3mm}
\begin{adjustwidth}{-1cm}{-1cm}
\renewcommand\arraystretch{1}
{\tabcolsep0.04in
\begin{tabular}{*{3}{c}*{13}{r}}
\midrule
Segment&&&\multicolumn{2}{c}{I}&\multicolumn{4}{c}{II}&\multicolumn{2}{c}{III}&\multicolumn{5}{c}{IV}\\
\cmidrule(lr){4-5}\cmidrule(lr){6-9}\cmidrule(lr){10-11}\cmidrule(lr){12-16}
Scenarios&$n$&&\multicolumn{1}{c}{$\hat{\alpha}_{1,1}$}&\multicolumn{1}{c}{$\hat{\gamma}_1$}&\multicolumn{1}{c}{$\hat{\alpha}_{1,2}$}&\multicolumn{1}{c}{$\hat{\alpha}_{2,2}$}&\multicolumn{1}{c}{$\hat{\alpha}_{3,2}$}&\multicolumn{1}{c}{$\hat{\gamma}_2$}&\multicolumn{1}{c}{$\hat{\alpha}_{1,3}$}&\multicolumn{1}{c}{$\hat{\gamma}_3$}&\multicolumn{1}{c}{$\hat{\alpha}_{1,4}$}&\multicolumn{1}{c}{$\hat{\alpha}_{2,4}$}&\multicolumn{1}{c}{$\hat{\alpha}_{3,4}$}&\multicolumn{1}{c}{$\hat{\alpha}_{4,4}$}&\multicolumn{1}{c}{$\hat{\gamma}_4$}\\
\midrule
MCP-BiINAR$_1$&500&Bias&$-$0.0132 &0.0067 &0.0083 &0.0143 &$-$0.0718 &0.2704 &&&&&&&\\
($m^0=1$)&&MSE&0.0045 &0.0051 &0.0064 &0.0031 &0.0220 &0.2946 &&&&&&&\\
&1000&Bias&$-$0.0070 &0.0043 &$-$0.0042 &0.0108 &$-$0.0149 &0.0743 &&&&&&&\\
&&MSE&0.0022 &0.0027 &0.0024 &0.0013 &0.0031 &0.0529 &&&&&&&\\
&1500&Bias&$-$0.0059 &0.0034 &$-$0.0007 &0.0071 &$-$0.0100 &0.0471 &&&&&&&\\
&&MSE&0.0014 &0.0017 &0.0015 &0.0009 &0.0015 &0.0300 &&&&&&&\\
\cmidrule(lr){2-16}
MCP-BiINAR$_2$&500&Bias&$-$0.0214 &0.0146 &$-$0.0053 &$-$0.0130 &$-$0.0412 &0.2698 &$-$0.0135 &0.0374 &&&&&\\
($m^0=2$)&&MSE&0.0073 &0.0090 &0.0050 &0.0047 &0.0074 &0.3772 &0.0049 &0.0621 &&&&&\\
&1000&Bias&$-$0.0140 &0.0099 &$-$0.0010 &$-$0.0082 &$-$0.0224 &0.1297 &$-$0.0065 &0.0213 &&&&&\\
&&MSE&0.0038 &0.0045 &0.0026 &0.0025 &0.0033 &0.1240 &0.0023 &0.0293 &&&&&\\
&1500&Bias&$-$0.0072 &0.0031 &$-$0.0005 &$-$0.0031 &$-$0.0149 &0.0659 &$-$0.0035 &0.0100 &&&&&\\
&&MSE&0.0022 &0.0029 &0.0016 &0.0017 &0.0020 &0.0630 &0.0015 &0.0201 &&&&&\\
\cmidrule(lr){2-16}
MCP-BiINAR$_3$&500&Bias&$-$0.0217 &0.0173 &0.0004 &$-$0.0467 &$-$0.0941 &0.7004 &$-$0.0201 &0.0280 &0.1997 &0.0368 &$-$0.0589 &$-$0.1911 &$-$0.1108 \\
($m^0=3$)&&MSE&0.0079 &0.0094 &0.0118 &0.0164 &0.0364 &1.4724 &0.0072 &0.0568 &0.0473 &0.0070 &0.0099 &0.0392 &1.1892 \\
&1000&Bias&$-$0.0097 &0.0039 &0.0021 &$-$0.0238 &$-$0.0302 &0.2197 &$-$0.0091 &0.0200 &0.1447 &0.0231 &$-$0.0265 &$-$0.1522 &0.0171 \\
&&MSE&0.0032 &0.0039 &0.0053 &0.0051 &0.0071 &0.2079 &0.0032 &0.0282 &0.0262 &0.0038 &0.0041 &0.0302 &0.9130 \\
&1500&Bias&$-$0.0066 &0.0013 &$-$0.0009 &$-$0.0147 &$-$0.0209 &0.1628 &$-$0.0071 &0.0142 &0.1110 &0.0180 &$-$0.0192 &$-$0.1138 &$-$0.0500 \\
&&MSE&0.0022 &0.0028 &0.0033 &0.0033 &0.0037 &0.1206 &0.0024 &0.0199 &0.0169 &0.0027 &0.0029 &0.0209 &0.7092 \\
\midrule
MCP-NBINAR$_1$&500&Bias&$-$0.0280 &0.0125 &$-$0.0158 &0.0165 &$-$0.0271 &0.1795 &&&&&&&\\
($m^0=1$)&&MSE&0.0075 &0.0079 &0.0057 &0.0026 &0.0074 &0.1591 &&&&&&&\\
&1000&Bias&$-$0.0136 &0.0070 &$-$0.0089 &0.0126 &$-$0.0118 &0.0817 &&&&&&&\\
&&MSE&0.0039 &0.0036 &0.0026 &0.0016 &0.0026 &0.0551 &&&&&&&\\
&1500&Bias&$-$0.0082 &0.0035 &$-$0.0040 &0.0076 &$-$0.0091 &0.0667 &&&&&&&\\
&&MSE&0.0025 &0.0025 &0.0018 &0.0011 &0.0017 &0.0344 &&&&&&&\\
\cmidrule(lr){2-16}
MCP-NBINAR$_2$&500&Bias&$-$0.0427 &0.0165 &$-$0.0148 &$-$0.0207 &$-$0.0534 &0.3983 &$-$0.0135 &0.0298 &&&&&\\
($m^0=2$)&&MSE&0.0127 &0.0126 &0.0058 &0.0048 &0.0104 &0.4833 &0.0056 &0.0894 &&&&&\\
&1000&Bias&$-$0.0207 &0.0110 &$-$0.0113 &$-$0.0100 &$-$0.0235 &0.1837 &$-$0.0060 &0.0087 &&&&&\\
&&MSE&0.0058 &0.0063 &0.0033 &0.0026 &0.0042 &0.1440 &0.0029 &0.0420 &&&&&\\
&1500&Bias&$-$0.0156 &0.0041 &$-$0.0033 &$-$0.0069 &$-$0.0204 &0.1161 &$-$0.0033 &0.0021 &&&&&\\
&&MSE&0.0042 &0.0041 &0.0018 &0.0017 &0.0030 &0.0866 &0.0019 &0.0296 &&&&&\\
\cmidrule(lr){2-16}
MCP-NBINAR$_3$&500&Bias&$-$0.0550 &0.0320 &$-$0.0367 &$-$0.0395 &$-$0.0986 &0.9187 &$-$0.0467 &0.0400 &0.0623 &$-$0.0143 &$-$0.0594 &$-$0.1151 &1.0320 \\
($m^0=3$)&&MSE&0.0137 &0.0138 &0.0203 &0.0173 &0.0226 &1.6163 &0.0157 &0.0711 &0.0106 &0.0090 &0.0148 &0.0204 &3.0663 \\
&1000&Bias&$-$0.0191 &0.0069 &$-$0.0020 &$-$0.0031 &$-$0.0635 &0.3124 &$-$0.0180 &0.0104 &0.0432 &$-$0.0078 &$-$0.0267 &$-$0.0737 &0.5681 \\
&&MSE&0.0062 &0.0059 &0.0083 &0.0093 &0.0159 &0.3099 &0.0060 &0.0053 &0.0057 &0.0047 &0.0052 &0.0135 &1.1605 \\
&1500&Bias&$-$0.0179 &0.0071 &$-$0.0035 &$-$0.0008 &$-$0.0407 &0.1833 &$-$0.0148 &0.0073 &0.0325 &$-$0.0034 &$-$0.0212 &$-$0.0507 &0.3659 \\
&&MSE&0.0043 &0.0039 &0.0051 &0.0055 &0.0104 &0.1270 &0.0043 &0.0033 &0.0040 &0.0032 &0.0033 &0.0094 &0.6941 \\
\midrule
\end{tabular}
}
\end{adjustwidth}
\end{table}

\section{Real data}\label{Set6}
In this section, we apply the proposed Auto-PGINARM-SA procedure to identify changes in three examples of real data series.
\subsection{Poilo data}
\begin{figure}[ht]
\begin{adjustwidth}{-2cm}{-1cm}
\includegraphics[width=8in,height=2.8in]{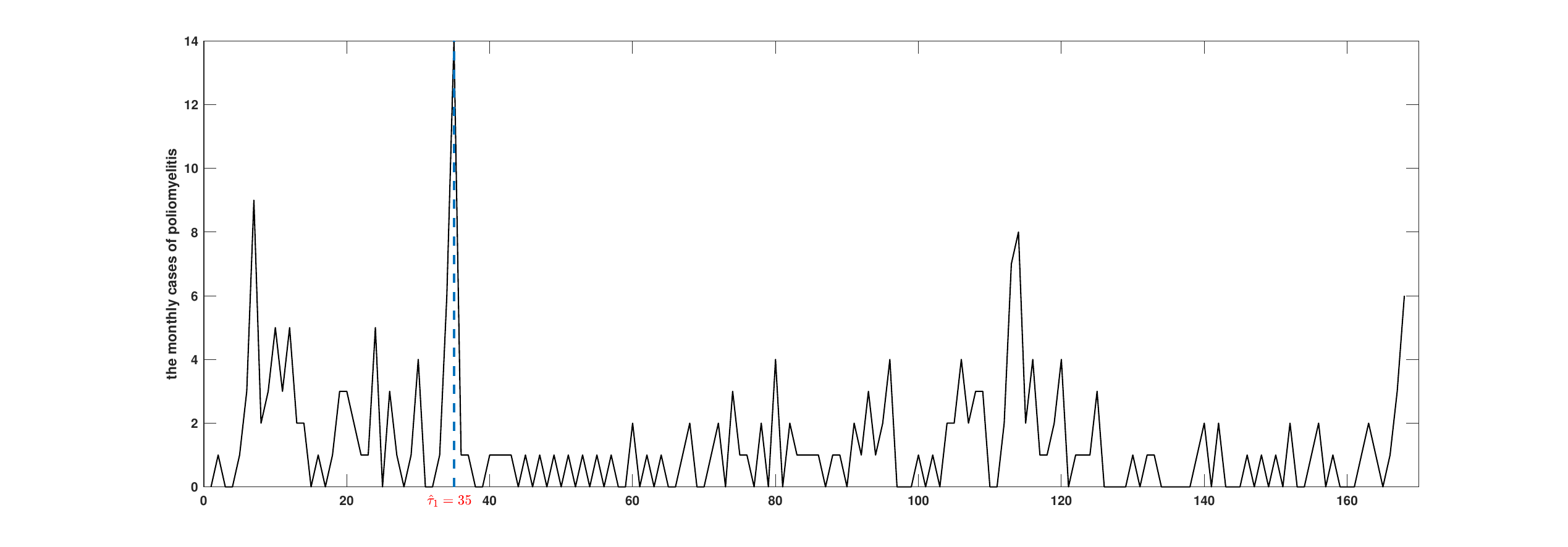}
\end{adjustwidth}
\vspace{-5mm}
\caption{Sample path of the monthly cases of poliomyelitis data.\protect\\ The dotted line represents the estimated location of the  change-point $\hat{\tau}_1$.}
\label{poilo}
\end{figure}

Firstly, we consider the monthly cases of poliomyelitis data in the US for a period of 14 years from 1970 to 1983 (see Figure \ref{poilo}).
There are 168 observations reported by the Centers for Disease Control and Prevention and have already been analyzed by several authors.
By investigating the autocorrelation function (ACF) and partial ACF plots from the data and by observing the spikes, \cite{Zheng2006} proposed that RCINAR(1) model could be used to fit this set of data. Furthermore, \cite{Kang2009} demonstrated by cumulative sum (CUSUM) that there exist change-point in these data, and obtained that the change-point location was the $35$th observation point through the position where the statistics reached the maximum value.

In this subsection, we apply the Auto-PGINARM-SA procedure to determine the location of the change-point and the corresponding parameters, and finally get the following MCP-GINAR estimated model,
\begin{align*}
{\rm E}(X_{t}|\mathcal{F}_{t-1})
\left\{
\begin{array}{lr}
0.2332X_{t-1,1}+1.8183,&0<t\leq 35,\\
(0.2279)~~~~~~~~(0.4051)&\\
0.2855X_{t-1,2}+0.7574,&36\leq t\leq 168.\\
(0.1313)~~~~~~~~(0.1111)&\\
\end{array}\right.
\end{align*}
where in parentheses are the standard errors (SE) of the estimators obtained from the square roots of the elements in the diagonal of the matrix $\hat{\bm{\Sigma_j}}=\bm{\hat{J}_j^{-1}}\bm{\hat{I}_j}\bm{\hat{J}_j^{-1}}$ computed on each segment $j$, with $\bm{\hat{J}_j}$ and $\bm{\hat{I}_j}$ are given by (\ref{SE1}) - (\ref{SE2}). The results show that there is a change-point in this data set, and the change-point location is $\hat{\tau}_1=35$ (see Figure \ref{poilo}). This is consistent with the results of \cite{Kang2009}.
\subsection{Technofirst stock trades data}
Secondly, we consider the daily number of trades in the stock of Technofirst listed in the NYSE Euronext group. This data set consists of 1662 observations from August 13, 2003 to December 3, 2013, and is available on the website ``\url{https://www.euronext.com/en/products/equities/FR0011651819-ALXP}".
Also, these data have been analyzed by many scholars. For example, \cite{Ahmad Francq2016} analyzed six stocks data listed in the NYSE Euronext group, including CR.FONC.MONACO, Siraga, Technofirst, Siparex Croissance, Proximedia and Achmea.
From their times series and ACF figures, it is not difficult to see that the stock trading data of Siraga, Technofirst, Siparex Crolssance and Achmea are all poissble to have change-points.
Furthermore, \cite{DiopKengne2021b} analyzed part of the data in Technofirst (corresponding to 1020-1662 of the data analyzed by us) and confirmed the existence of change-points. They got the locations of change-points are 230 and 311 (corresponding to 1249 and 1330 in our data).

\begin{figure}[H]
\begin{adjustwidth}{-2cm}{-1cm}
\includegraphics[width=8in,height=2.8in]{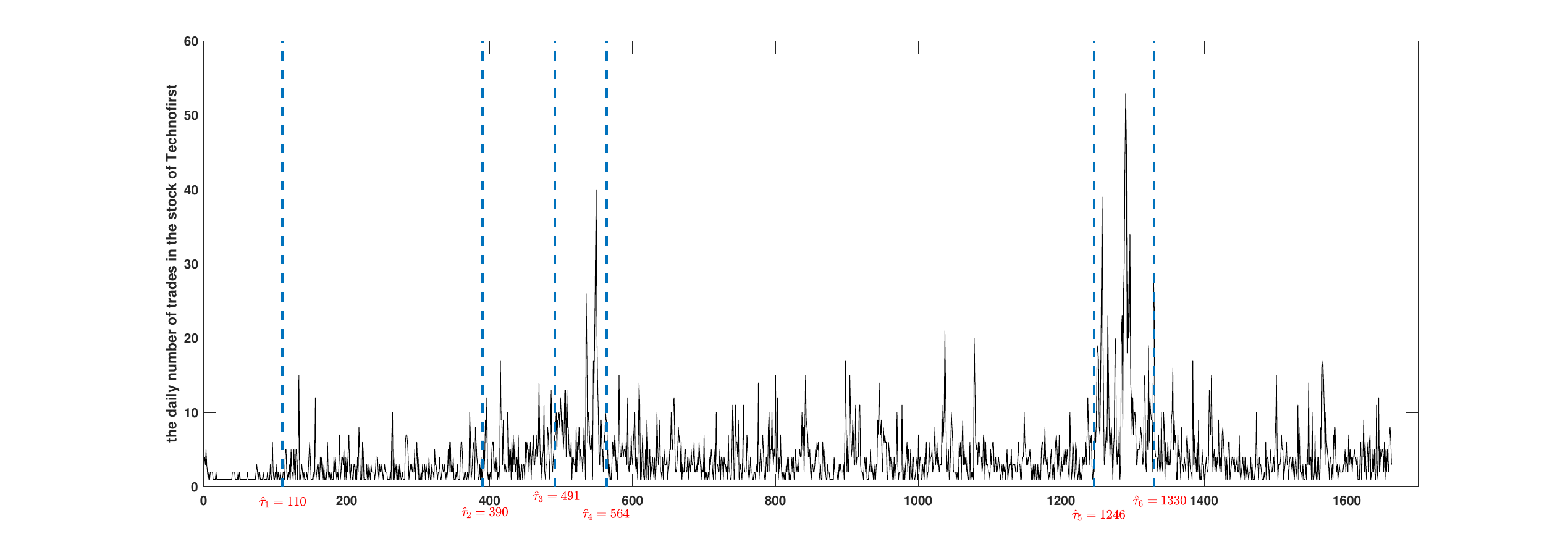}
\end{adjustwidth}
\vspace{-5mm}
\caption{Sample path of the daily number of trades in the stock of Technofirst data.\protect\\ The dotted lines represent the estimated locations of the change-points $\hat{\tau}_1$ - $\hat{\tau}_6$.}
\label{Technofirst}
\end{figure}

In this subsection, we also apply the proposed Auto-PGINARM-SA procedure to this financial time series data. Figure \ref{Technofirst} shows the time series and the estimated locations of change-points. Through the Auto-PGINARM-SA procedure, the number of change-points $\hat{m}=6$, the change-points locations $\hat{\bm{\tau}}=\{110, 390, 491, 564, 1246, 1330\}$ are obtained, and the corresponding estimated model is as follows:
\begin{align*}\footnotesize
{\rm E}(X_{t}|\mathcal{F}_{t-1})
=\left\{
\begin{array}{lr}
0.2922X_{t-1,1}+1.0038,&0<t\leq 110,\\
(0.1096)~~~~~~~~~~(0.1632)&\\
0.1572X_{t-1,2}+2.1611,&111\leq t\leq 390,\\
(0.0746)~~~~~~~~~~(0.1880)&\\
0.1523X_{t-1,3}+3.7013,&391\leq t\leq 491,\\
(0.0874)~~~~~~~~~~(0.4299)&\\
0.6099X_{t-1,4}+3.2385,&492\leq t\leq 564,\\
(0.1902)~~~~~~~~~~(1.7358)&\\
0.3161X_{t-1,5}+2.5951,&565\leq t\leq 1246,\\
(0.0438)~~~~~~~~~~(0.1687)&\\
0.6661X_{t-1,6}+4.2140,&1247\leq t\leq 1330,\\
(0.1124)~~~~~~~~~~(1.3606)&\\
0.2182X_{t-1,7}+3.0588,&1331\leq t\leq 1662.\\
(0.0732)~~~~~~~~~~(0.2917)&\\
\end{array}\right.
\end{align*}
It can be seen that the change-points locations 1246 and 1330 obtained by using the Auto-PGINARM-SA procedure are consistent with 1249 and 1330 obtained by \cite{DiopKengne2021b}. This shows that our proposed model and method can also effectively analyze the locations of change-points. However, the SE of the estimators in segments 4 and 6 were observed to be insignificant, some other models, such as INGARCH models, should be considered to fit the data in these two segments in future studies.
\subsection{Siraga stock trades data}
\begin{figure}[H]
\begin{adjustwidth}{-2cm}{-1cm}
\includegraphics[width=8in,height=2.8in]{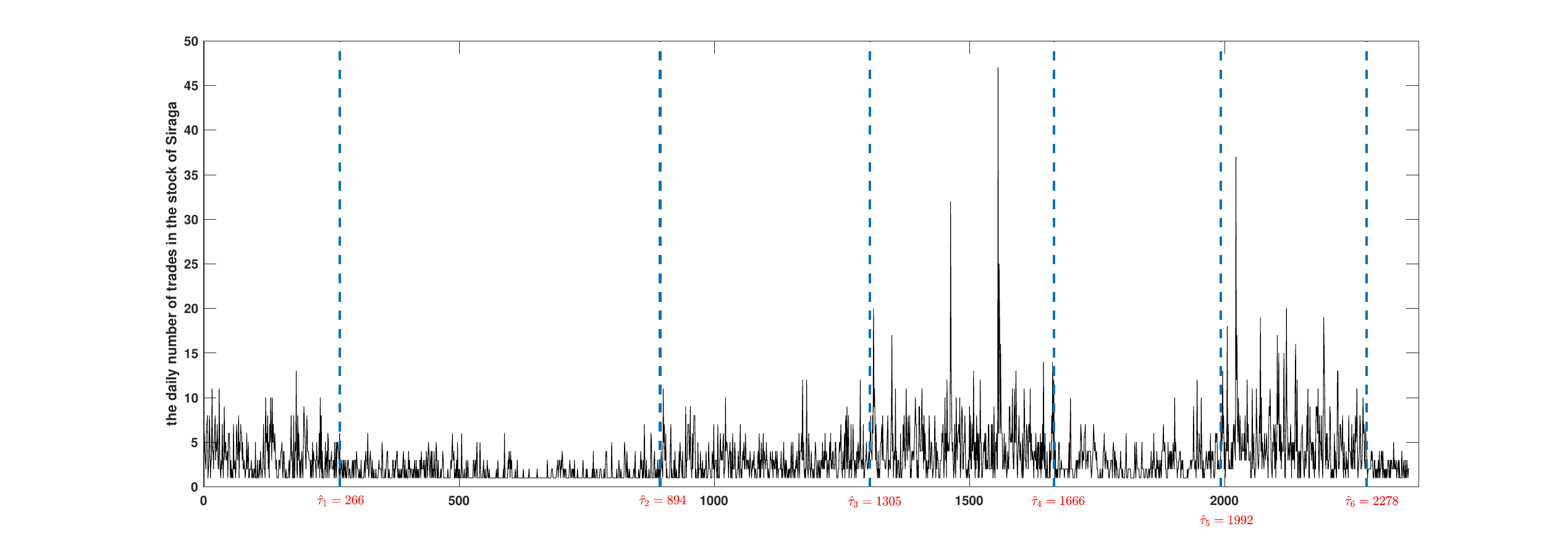}
\end{adjustwidth}
\vspace{-5mm}
\caption{Sample path of the daily number of trades in the stock of Siraga data.\protect\\ The dotted lines represent the estimated locations of the change-points $\hat{\tau}_1$ - $\hat{\tau}_6$.}
\label{Siraga}
\end{figure}

In this subsection, we consider another financial time series data, that is, the daily number of trades in the stock of Siraga listed in the NYSE Euronext group. This data set consists of 2362 observations and is available on the website ``\url{https://live.euronext.com/en/product/equities/FR0000060170-XPAR}".
Similar to the previous two applications, Figure \ref{Siraga} shows the time series and the estimated locations of change-points. Through the Auto-PGINARM-SA procedure, we obtained the number of change-points $\hat{m}=6$, the change-points locations $\hat{\bm{\tau}}=\{266, 894, 1305, 1992, 2278, 2362\}$, and the corresponding estimated model is as follows:
\begin{align*}
\scriptsize
{\rm E}(X_{t}|\mathcal{F}_{t-1})
=\left\{
\begin{array}{lr}
0.0610X_{t-1,1}+0.1144X_{t-2,1}+0.0001X_{t-3,1}+0.1273X_{t-4,1}+0.2038X_{t-5,1}+1.7422,&0<t\leq 266,\\
(0.0538)~~~~~~~~~(0.0611)~~~~~~~~~~~(0.0637)~~~~~~~~~~~(0.0635)~~~~~~~~~~(0.0658)~~~~~~~~~~(0.3327)&\\
0.2061X_{t-1,2}+1.3324,&267\leq t\leq 894,\\				
(0.0438)~~~~~~~~~(0.0792)\\ 				
0.0998X_{t-1,3}+2.5934,&895\leq t\leq 1305,\\				
(0.0495)~~~~~~~~~(0.1573)&\\				
0.1234X_{t-1,4}+0.2269X_{t-2,4}+3.0545,&1306<t\leq 1666,\\			
(0.0540)~~~~~~~~~(0.0617)~~~~~~~~~~~(0.4524)&\\ 			
0.3352X_{t-1,5}+1.7640,&1667\leq t\leq 1992,\\				
(0.0609)~~~~~~~~~(0.1554)&\\ 				
0.2007X_{t-1,6}+4.0554,&1993\leq t\leq 2278,\\				
(0.0537)~~~~~~~~~(0.3484)&\\ 				
0.0001X_{t-1,7}+2.2458,&2279\leq t\leq 2362.\\				
(0.1041)~~~~~~~~~(0.2359)&\\ 	
\end{array}\right.
\end{align*}
Clearly, all estimators of the estimated MCP-GINAR model  except the last segment are significant regarding the standard errors.
For the last segment, judging from the results, it seems more appropriate to fit it with the integer-valued moving average (INMA) model.

\section{Conclusion and discussion}\label{Set7}
This paper introduces a piecewise stationary model, MCP-GINAR, to describe the common non-stationary processes in statistics.
The MDL criterion based on the PQML for the MCP-GINAR model has been proposed to estimate the number of change-points, the location of the change-points, the order of each segment and its parameters.
Regarding the consistency of the MDL model selection procedure, we reach the following conclusions.
When the number of change-points is known, if ${\rm E}|X_{t,j}|^{2+\eta}<\infty$, estimators based on MDL procedure are strongly consistent.
Under the condition of the number of change-points is unknown, if ${\rm E}|X_{t,j}|^{4+\eta}<\infty$,
estimators are weakly consistent, further, if ${\rm E}|X_{t,j}|^{8+\eta}<\infty$, estimators are strongly consistent.
Regarding optimizing MDL, an improved genetic algorithm with simulated annealing, Auto-PGINARM-SA, is proposed.
The simulation results in Section \ref{Set5} show that the Auto-PGINARM-SA program is very satisfactory.
It can not only find solutions, but also effectively solve the precocious problem of the genetic algorithm.
Finally, we use Auto-PGINARM-SA to complete the change-points estimation analysis for three applications, and the analysis results indicate that the proposed procedure is useful in practice.

However, we must point out that there are many topics that are worth further research.
On the one hand, the extension of the MCP-GINAR model will be an important issue. As the results of the second and third applications in Section \ref{Set6} show, although the Auto-PGINARM-SA procedure can effectively solve the change-points estimation problem, some other models, such as INGARCH models, INMA models, may be more suitable to fit the data. So extending the MCP-GINAR(${\bm{p}}$) model to MCP-INGARCH(${\bm{p,q}}$) will be an interesting topic.
On the other hand, the optimization of MDL will be a popular topic for MCP-GINAR, such as Dynamic Programming, Optimal Partitioning, Pruned Exact Linear Time algorithms are all worth trying.
We will leave the above issues as our future work.

\section*{Acknowledgements}
This work is supported by National Natural Science Foundation of China (No.12271231, 12001229, 11901053).
\section*{Appendix}
As stated in Section \ref{Set3}, in practice, the observed likelihood $\tilde{L}_{n_j}^{(j)}(\bm{\theta_j};p_j,\bm{x_j})$ is used to approximate the true likelihood $L_{n_j}^{(j)}(\bm{\theta_j};p_j,\bm{x_j})$, so let us first state the following Lemma to control this approximation. Also, this Lemma will be used to prove all theorems.
\begin{lemma}\label{lemma1}
Let $\{X_t\}$ be a piecewise stationary process MCP-GINAR process defined in (\ref{CGINAR}). If ${\rm E}(X_{t,j})^{2k+\eta}<\infty$ for some $\eta>0$, then the Assumption 1 ($k$) in \cite{DavisYau2013} can be verfied. That is, for any $j=1,2,...,m+1$ and fixed $p_j$, the function $\ell_t^{(j)}$ is two-time continuously differentiable with respective to $\bm{\theta_j}$ and the first and second derivatives $L_{n_j}^{'(j)}, \tilde{L}_{n}^{'(j)}$ and $L_{n}^{{''}(j)}, \tilde{L}_{n}^{{''}(j)}$, respectively, of the function defined in (\ref{PQMLeq}) and (\ref{tilde_L}), satisfy
\begin{align}
\sup\limits_{\bm{\theta_j}\in\Theta_j(p_j)}|\frac{1}{n}L_{n_j}^{(j)}(\bm{\theta_j};p_j,\bm{x_j})-\frac{1}{n}\tilde{L}_{n_j}^{(j)}(\bm{\theta_j};p_j,\bm{x_j})|&=o(n^{\frac{1}{k}-1}),\label{assump1a}\\
\sup\limits_{\bm{\theta_j}\in\Theta_j(p_j)}|\frac{1}{n}L_{n_j}^{'(j)}(\bm{\theta_j};p_j,\bm{x_j})-\frac{1}{n}\tilde{L}_{n_j}^{'(j)}(\bm{\theta_j};p_j,\bm{x_j})|&=o(n^{\frac{1}{k}-1}),\label{assump1b}\\
\sup\limits_{\bm{\theta_j}\in\Theta_j(p_j)}|\frac{1}{n}L_{n_j}^{''(j)}(\bm{\theta_j};p_j,\bm{x_j})-\frac{1}{n}\tilde{L}_{n_j}^{''(j)}(\bm{\theta_j};p_j,\bm{x_j})|&=o(1),\label{assump1c}
\end{align}
almost surely.
\end{lemma}
$\mathbf{Proof~of~Lemma~1.}$
We first prove (\ref{assump1a}),
\begin{align*}
&|L_{n_j}^{(j)}(\bm{\theta_j};p_j,\bm{x_j})-\tilde{L}_{n_j}^{(j)}(\bm{\theta_j};p_j,\bm{x_j})|\\
&=\left|\sum\limits_{t=1}^{n_j}X_{t,j}\log\left(\frac{\xi_{t,j}(\bm{\theta_j};p_j|X_{s,j},s<t)}{\xi_{t,j}(\bm{\theta_j};p_j|\bm{\tilde{x}_{i,j}})}\right)
-\left[\xi_{t,j}(\bm{\theta_j};p_j|X_{s,j},s<t)-\xi_{t,j}(\bm{\theta_j};p_j|\bm{\tilde{x}_{i,j}})\right]\right|\nonumber\\
&\leq\sum\limits_{t=1}^{n_j}\left(\left|X_{t,j}\log(\upsilon_{t,j}^{(1)})\right|+\left|\upsilon_{t,j}^{(2)}\right|\right),\nonumber
\end{align*}
where
\begin{align*}
\upsilon_{t,j}^{(1)}=\frac{\sum\limits_{k=1}^{p_j}\alpha_{k,j} X_{t-k,j}+\gamma_j}{\sum\limits_{k=1}^{p_j}\alpha_{k,j} X_{\tau_{j-1}+t-k}+\gamma_j},~~~\upsilon_{t,j}^{(2)}=\sum\limits_{k=1}^{p_j}\alpha_{k,j} (X_{t-k,j}-X_{\tau_{j-1}+t-k}).
\end{align*}
Since $\alpha_{k,j}\in(0,1)$ and $\gamma_j>0$, there exist a suitable constant $0<C_1<\infty$ satisfy $|\log(\upsilon_{t,j}^{(1)})|\leq C_1X_{t,j}$, thus,
\begin{align*}
|L_{n_j}^{(j)}(\bm{\theta_j};p_j,\bm{x_j})-\tilde{L}_{n_j}^{(j)}(\bm{\theta_j};p_j,\bm{x_j})|\leq\sum\limits_{t=1}^{n_j}\left(C_1|X_{t,j}|^2+\left|\upsilon_{t,j}^{(2)}\right|\right)\nonumber.
\end{align*}
Note that when ${\rm E}(X_{t,j})^{2k}<\infty$, from Borel-Cantelli lemma, for any $K>0$,
\begin{align*}
\sum\limits_{t=1}^{n_j}P(X_{t,j}^2>Kn^{\frac{1}{k}})=\sum\limits_{t=1}^{n_j}P(X_{t,j}^{2k}>K^{k}n)\leq\frac{1}{K^{k}}{\rm E}(X_t,j)^{2k}<\infty,\\
\sum\limits_{t=1}^{n_j}P(|\upsilon_{t,j}^{(2)}|>Kn^{\frac{1}{k}})=\sum\limits_{t=1}^{n_j}P(|\upsilon_{t,j}^{(2)}|^{k}>K^{k}n)\leq\frac{1}{K^{k}}{\rm E}|\upsilon_{t,j}^{(2)}|^{k}<\infty,
\end{align*}
thus when ${\rm E}(X_{t,j})^{2k+\eta}<\infty$, we have $\sum\limits_{t=1}^{n_j}\left(C_1|X_{t,j}|^2+\left|\upsilon_{t,j}^{(2)}\right|\right)=o(n^{\frac{1}{k}})$ almost surely.
In the similar way, (\ref{assump1b}) and (\ref{assump1c}) hold if ${\rm E}(X_{t,j})^{2k+\eta}<\infty$. The proof of Lemma \ref{lemma1} is completed.

Next, we use the following Lemma to illustrate some conditions of PQML function, which are also conditions for the consistency and asymptotic normality of the PQML estimator in \cite{Ahmad Francq2016}.
\begin{lemma}\label{lemma2}
Let $\{X_t\}$ be a piecewise stationary MCP-GINAR process defined in (\ref{CGINAR}). If ${\rm E}(X_{t,j})^{2k+\eta}<\infty$ for some $\eta>0$, then the Assumption 2 ($k$) in \cite{DavisYau2013} can be verfied. That is, for any $j=1,2,...,m+1$ and fixed $p_j$, there exists an $\epsilon>0$ such that
\begin{align}
\sup\limits_{\bm{\theta_j}\in\Theta_j(p_j)}{\rm E}|\ell_t^{(j)}(\bm{\theta_j};p_j, X_{t,j}|X_{s,j},s<t)|^{k+\epsilon}&<\infty,\label{lemma2.1}\\
\sup\limits_{\bm{\theta_j}\in\Theta_j(p_j)}{\rm E}|\ell_t^{'(j)}(\bm{\theta_j};p_j, X_{t,j}|X_{s,j},s<t)|^{k+\epsilon}&<\infty,\label{lemma2.2}\\
\sup\limits_{\bm{\theta_j}\in\Theta_j(p_j)}{\rm E}|\ell_t^{''(j)}(\bm{\theta_j};p_j, X_{t,j}|X_{s,j},s<t)|&<\infty,\label{lemma2.3}
\end{align}
where $\ell_t^{'(j)}$ and $\ell_t^{''(j)}$ are the first and second derivatives of $\ell_t^{(j)}$, $\ell_t^{(j)}(\bm{\theta_j};p_j, X_{t,j}|X_{s,j},s<t)$ is defined in (\ref{PQMLeq}).
\end{lemma}
$\mathbf{Proof~of~Lemma~2.}$ $\ell_t^{(j)}$ and its first and second derivatives are given by
\begin{align*}
&\ell_t^{(j)}(\bm{\theta_j};p_j, X_{t,j}|X_{s,j},s<t)=X_{t,j}\log \xi_{t,j}(\bm{\theta_j};p_j|X_{s,j},s<t)-\xi_{t,j}(\bm{\theta_j};p_j,|X_{s,j},s<t),\\
&\ell_t^{'(j)}(\bm{\theta_j};p_j|X_{s,j},s<t)=\frac{1}{\xi_{t,j}(\bm{\theta_j})}\frac{\partial\xi_{t,j}(\bm{\theta_j})}{\partial \bm{\theta_j}}\frac{\partial\xi_{t,j}(\bm{\theta_j})}{\partial \bm{\theta_j}^{\mathrm{T}}},\\
&\ell_t^{''(j)}(\bm{\theta_j};p_j|X_{s,j},s<t)=(\frac{X_{t,j}}{\xi_{t,j}(\bm{\theta_j})}-1)^2\frac{\partial\xi_{t,j}(\bm{\theta_j})}{\partial \bm{\theta_j}}\frac{\partial\xi_{t,j}(\bm{\theta_j})}{\partial \bm{\theta_j}^{\mathrm{T}}}.
\end{align*}
By the same arguments as the Lemma \ref{lemma1}, there exist a suitable constant $0<C_2<\infty$ satisfy ${\rm E}|\ell_t^{(j)}(\bm{\theta_j};p_j, X_{t,j}|X_{s,j},s<t)|\leq C_2{\rm E}(X_{t,j})^{2}$, thus (\ref{lemma2.1}) holds if ${\rm E}(X_{t,j})^{2k+\eta}<\infty$.
On the other hand, according to the arguments in Section 3.7. of \cite{Ahmad Francq2016}, we have
$$\ell_t^{'(j)}(\bm{\theta_j};p_j, X_{t,j}|X_{s,j},s<t)\leq C_3+C_4 X_{t,j},$$
$$\ell_t^{''(j)}(\bm{\theta_j};p_j, X_{t,j}|X_{s,j},s<t)\leq C_5(C_3+C_4 X_{t,j}),$$
where $C_3,C_4,C_5$ are some suitable constants. Thus if ${\rm E}(X_{t,j})^{2k+\eta}<\infty$, then (\ref{lemma2.2}) and (\ref{lemma2.3}) can be verified.
~\\
$\mathbf{Proof~of~Theorem~\ref{MDL_mknown_theorem}.}$ In order to prove Theorem~\ref{MDL_mknown_theorem}, it is sufficient to show that Assumptions 1 (1), 2 (1) and 3 of \cite{DavisYau2013} holds. Note that if we take $k=1$, that is  ${\rm E}(X_{t,j})^{2+\eta}<\infty$, Assumptions 1 (1), 2 (1) are satisfied according to Lemma \ref{lemma1} and \ref{lemma2}. Assumption 3 can be verified by the ergodic theorem and the compactness of the parameter space. The proof of Theorem \ref{MDL_mknown_theorem} is completed.
~\\
$\mathbf{Proof~of~Theorem~\ref{week}.}$ For Theorem~\ref{week} to hold we verify Assumptions 1 (1), 1*, 2 (2), 3 and 5 in \cite{DavisYau2013}. Similar to the argument in the proof of Theorem~\ref{MDL_mknown_theorem}, if the $(2*2+\eta)$-th moment of $X_{t,j}$ exists, then Assumptions 1 (1), 1*, 2 (2) and 3 hold. Next we verify that Assumption 5 holds. Considering that $\bm{\theta_j}$ belongs to the interior of $\Theta_j(p_j)$ Assumption 5A) is easily demonstrated for GINAR($p_j$) process according to PQML in {\color{blue}Ahmad and Francq (2016)}. On the other hand, GINAR($p_j$) can be written as an AR($p_j$) process according to Remark \ref{remark2}.
Similar to the arguments of Example 2 in \cite{DavisYau2013}, we can verify that Assumption 5B) holds. Thus the proof of Theorem \ref{week} is completed.
~\\
$\mathbf{Proof~of~Theorem~\ref{strong}.}$ For Theorem~\ref{strong} to hold we verify Assumptions 1 (2), 2 (4), 3 , 5, and either 4 (0.5) or 4* in \cite{DavisYau2013}. Similar to the argument in the proof of Theorem~\ref{week}, if the $(2*4+\eta)$-th moment of $X_{t,j}$ exists, then Assumptions 1 (2), 2 (4), 3 and 5 hold. Therefore, it remains to verify Assumption 4 (0.5) or 4*. It is well known that GINAR($p_j$) process is a stationary ergodic Markov chain according to Remark \ref{remark1}, that is, GINAR($p_j$) process is strongly mixing with geometric rate. Also, GINAR($p_j$) can be written as an AR($p_j$) process according to Remark \ref{remark2}. Thus $\ell_t^{(j)}(\bm{\theta_j};p_j, X_{t,j}|X_{s,j},s<t)$ is strongly mixing with the same geometric rate as it is a function of finite number
of the strongly mixing $X_{t,j}$'s (Theorem 14.1 of \cite{Davidson1994}). Thus Assumption 4* holds. In fact, Lemma 1 in \cite{DavisYau2013} already states that Assumption 4(0.5) holds under Assumption 2(2) and 4*. Note that Assumption 2(2) is obviously satisfied under ${\rm E}(X_{t,j})^{8+\eta}<\infty$. Therefore, Assumption 4(0.5) is also satisfied under our assumptions. The proof of Theorem \ref{strong} is completed.

\end{document}